\documentclass[9pt,twocolumn,twoside]{article}

\usepackage[paperwidth=210mm,paperheight=297mm,centering,hmargin=2cm,vmargin=2.0cm]{geometry}  
\usepackage{caption}
\usepackage{amsfonts}
\usepackage{amsmath}
\usepackage{amssymb}
\usepackage{graphicx}
\usepackage{siunitx} 
\usepackage{xcolor}
\usepackage{cite}
\usepackage{authblk} 
\usepackage{titlesec}

\usepackage{dcolumn}
\usepackage{bm}
\usepackage{epstopdf}
\usepackage{upgreek}  
\usepackage[normalem]{ulem} 
\newcommand\redsout{\bgroup\markoverwith{\textcolor{red}{\rule[0.5ex]{2pt}{1.0pt}}}\ULon}

\usepackage{float}
\usepackage{isomath} 
\usepackage{tensor}
\usepackage[inline]{enumitem} 

\titleformat{\subsubsection}[runin]{\bfseries}{}{}{}[.]

\renewcommand{\v}[1]{\ensuremath{\mathbf{#1}}} 
\newcommand{\gv}[1]{\ensuremath{\mbox{\boldmath$ #1 $}}}

\newcommand{\uv}[1]{\ensuremath{\mathbf{\hat{#1}}}} 
\newcommand{\abs}[1]{\left| #1 \right|} 
\renewcommand{\d}[2]{\frac{\mathrm{d} #1}{\mathrm{d} #2}} 

\renewcommand{\div}[1]{\gv{\nabla} \cdot #1} 
\let\baraccent=\= 
\renewcommand{\=}[1]{\stackrel{#1}{=}} 
\DeclareMathAlphabet{\mathsfsl}{OT1}{cmss}{m}{sl}

\title{\textbf{\LARGE{Polar state memory in active fluids}}}

\author[1]{\small{Bo Zhang}}
\author[2]{Hang Yuan}
\author[1]{Andrey Sokolov}
\author[3, 4]{Monica Olvera de la Cruz}
\author[1,*]{Alexey Snezhko}

\affil[1]{Materials Science Division, Argonne National Laboratory,
9700 South Cass Avenue, Argonne, IL 60439, USA.}
\affil[2]{Applied Physics Graduate Program, Northwestern University,
2145 Sheridan Road, Evanston, IL 60208, USA.}
\affil[3]{Department of Materials Science and Engineering, Northwestern University,
2220 Campus Drive, Evanston, IL 60208, USA.}
\affil[4]{Department of Physics and Astronomy, Northwestern University,
2145 Sheridan Road, Evanston, IL 60208, USA.}

\affil[*]{Corresponding author: snezhko@anl.gov}

\date{\small{\today}}

\begin{document}

\maketitle


\noindent
\textbf{
Spontaneous emergence of correlated states such as flocks and vortices are prime examples of remarkable collective dynamics and self-organization observed in active matter~\cite{vicsek2012collective, marchetti2013hydrodynamics,aranson2013active,snezhko2016complex,zottl2014hydrodynamics,wu2017transition}. The formation of globally correlated polar states in geometrically confined systems proceeds through the emergence of a macroscopic steadily rotating vortex that spontaneously selects a clockwise or counterclockwise global chiral state~\cite{bricard2015emergent,kaiser2017flocking}.
Here, we reveal that a global vortex formed by colloidal rollers exhibits state memory. The information remains stored even when the energy injection is ceased and the activity is terminated. We show that a subsequent formation of the collective states upon re-energizing the system is not random. We combine experiments and simulations to  elucidate how a combination of hydrodynamic and electrostatic interactions leads to hidden asymmetries in the local particle positional order encoding the chiral state of the system. The stored information can be accessed and exploited to systematically command subsequent polar states of active liquid through temporal control of the activity. With the chirality of the emergent collective states controlled on-demand, active liquids offer new possibilities for flow manipulation, transport, and mixing at the microscale.
}

\begin{figure*}[!t] 
\centering
\includegraphics[width=1.0\linewidth]{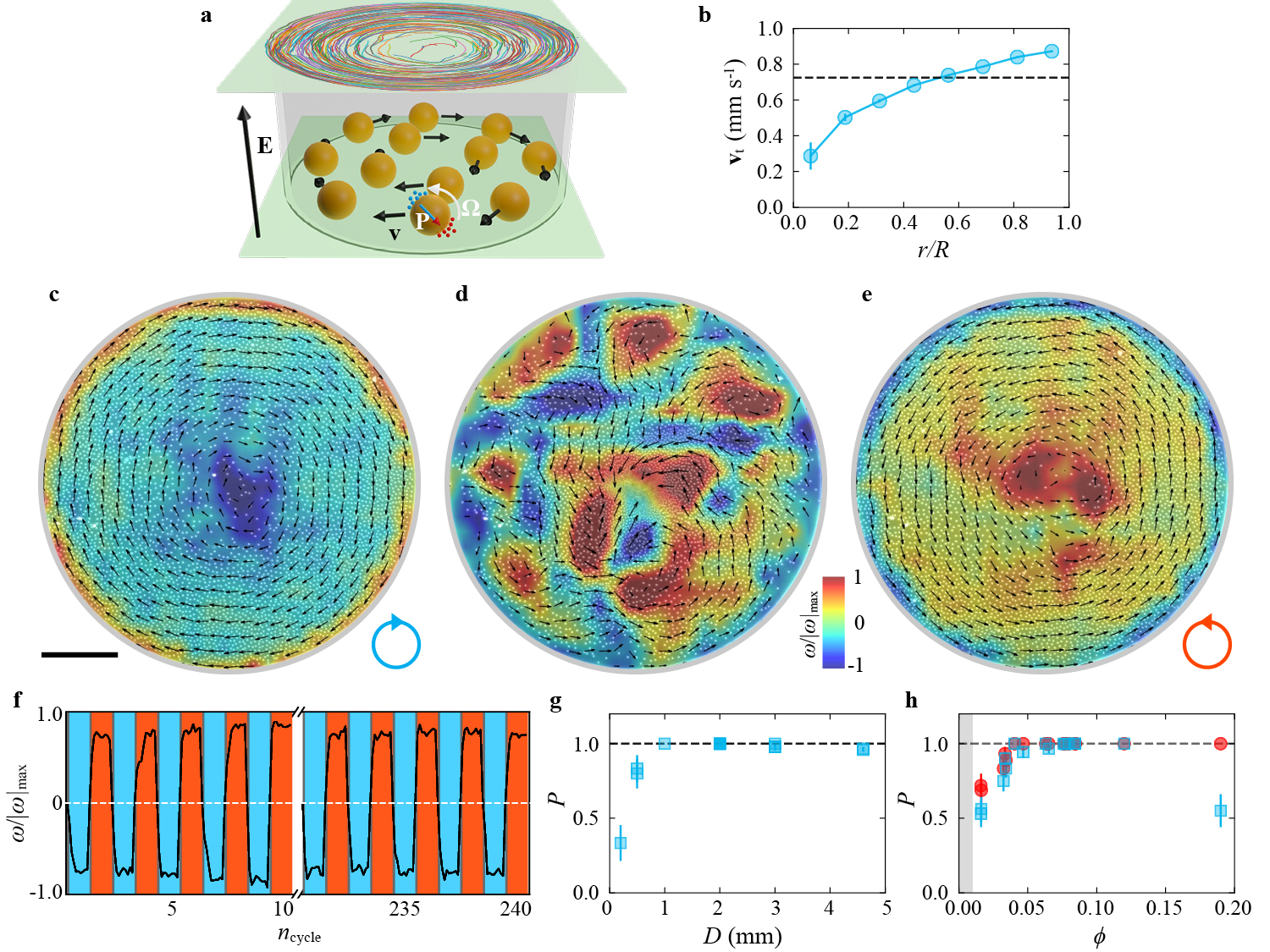}
\caption{
\textbf{Vortex chiral state reversal induced by a temporal modulation of activity.}
	\textbf{a}, A sketch of the experimental setup. Red and blue specks illustrate induced positive and negative charges around a particle. Particles develop a tilted dipole $\bf{P}$ and rotate with an angular velocity $\bf{\Omega}$ resulting in a translational velocity $\bf{v}$ along the surface.
Particle trajectories in a typical self-organized vortex of rollers are illustrated as an insert on the top cell surface. The local particle density in the vortex is not uniform with a noticeable depletion in the center of the well.
\textbf{b}, Averaged tangential velocities $v_\text{t}$ of particles in the roller vortex as a function of the distance from the center. $R$ is the radius of the well. The dash line shows the averaged velocity $v_0$ of an isolated roller measured in a dilute sample under the same experimental conditions. The error bars are standard deviations of the mean.
\textbf{c-e}, Snapshots of superimposed velocity (arrows) and vorticity (background color) fields of the rollers during a vortex reversal under a temporal modulation of the activity. \textbf{c}, A stable initial roller vortex with a clockwise rotation. \textbf{d}, Intermittent flocking after re-activation of the system. Rollers initially move mostly towards the center of the cell.
\textbf{e}, A stable vortex with a counter-clockwise rotation evolved from the state shown in \textbf{d}. The blue and red circles with an arrow illustrate the chirality of the states. The scale bar is 0.2 mm. The field strength $E=$ 2.7 V \SI{}{\micro\meter}$^{-1}$, the particle area fraction $\phi = 0.12$. See Supplementary Video 1.
	\textbf{f}, Cycling of the vortex chiral state by a temporal termination and subsequent restoration of the electric field. The black curve shows normalized magnitude of the vorticity in the vortex center. Blue spans indicate CW vortices, and red spans correspond to CCW vortices. The vortex reversal is observed for every one of 240 cycles. The particle area fraction $\phi = $ 0.08. See Supplementary Video 1 for details.
	\textbf{g}, The reversal probability $P$ as a function of the well diameter $D$.  The particle area fraction $\phi$ = 0.07$\pm$0.01. $P=N_\text{r}/N_\text{t}$, where $N_\text{r}$  is the number of times the vortex reversal is observed during $N_\text{t}$ activity cycles.
	\textbf{h}, The reversal probability $P$ (blue squares) as a function of the particles area fraction $\phi$ for $D =$ 2 mm. Red circles illustrate the probability of the vortex formation in the same system. A shaded area indicates the area fractions $\phi<$ 0.01 where no stable vortex is observed. Dash lines depict $P = 1$. Error bars are standard deviations of the mean.
}
\label{Fig1}
\end{figure*}

Ensembles of interacting self-propelled particles, or active matter, exhibit a plethora of remarkable collective phenomena which have been widely observed and studied in both biological and artificial systems \cite{peruani2012collective, schaller2010polar, sumino2012large, sokolov2012physical, bricard2013emergence, yan2016reconfiguring,  kokot2017active, doostmohammadi2018active, zhang2020reconfigurable}. Many synthetic active systems are realized by ensembles of externally energized particles~\cite{martin2013driving, driscoll2017unstable, weber2013long, kokot2018manipulation, massana2018active, han2020reconfigurable, palacci2013living}. The onset of globally correlated vortical states in active ensembles is associated with a spontaneous symmetry breaking between two equally probably chiral states (characterized by clockwise or counterclockwise rotations). Once the global state is formed, it is often robust and stable~\cite{bricard2015emergent,kaiser2017flocking}. Subsequent control of such polar states, however, remains elusive and largely unexplored.
Here we use a model system of colloidal rollers powered by the Quincke electro-rotation mechanism~\cite{quincke1896ueber, melcher1969electrohydrodynamics} to demonstrate an intrinsic chiral state memory in active liquids. The experimental system consists of polystyrene spheres dispersed in a weakly conductive liquid that are sandwiched between two ITO-coated glass slides and energized by a static (DC) electric field (Fig.~\ref{Fig1}a, see Methods for details). The particles continuously roll with a typical speed of $v_0 \sim $0.7 mm s$^{-1}$, while energized by a uniform DC electric field $E$ = 2.7 V \SI{}{\micro\meter}$^{-1}$. The rollers experience hydrodynamic and electrostatic interactions that promote alignment of their translational velocities \cite{bricard2013emergence}. At low particle number densities, the rollers move randomly and resemble the dynamics of an isotropic gas. Confined in a well at density above a certain threshold, the rollers self-organize into a single stable vortex (Fig.~\ref{Fig1}).
Trajectories of Quincke rollers in the vortex are nearly circular, and the average tangential velocity $\textbf{v}_\text{t}$ increases with the distance from the center (Fig.~\ref{Fig1}a,b).

The roller vortex is robust and remains stable as long as the system is energized.  Typical velocity and vorticity fields in a stable vortex are shown in Fig.~\ref{Fig1}c. Two possible chiral states of the vortex, clockwise (CW) or counter-clockwise (CCW), are equally probable, and the system  spontaneously selects one in the course of the vortex self-assembly from initially random distribution of particles.
When the electric field is switched off, particles come to rest within a time scale given by the Maxwell-Wagner relaxation time $\tau_\text{MW}$ and viscous time scale $\tau_{\nu}$ that are both of the order of 1 ms for our system. The particle arrangements appear to be random and the direction of the original vortex cannot be easily identified. A cessation of the activity  beyond $\tau_\text{MW}=(\epsilon_\text{p}+2\epsilon_\text{f})/(\sigma_\text{p}+2\sigma_\text{f})$, where $\epsilon_\text{p,f}$ and $\sigma_\text{p,f}$ are respective particle and fluid permittivities and conductivities, preserves the particle positions but erases their previous velocities, polarizations and resets aligning forces.
Once the system is re-energized by the same DC electric field, particles initially move mostly toward the center of the well where the density is the lowest (see Fig.~\ref{Fig1}d). Eventually the rollers redistribute over the well and form a single vortex with the direction of rotation opposite to the state preceding the cessation of the activity (Fig.~\ref{Fig1}e, see also Supplementary Video 1).

The probability of the vortex reversal, $P$, achieves 1 in a wide range of explored experimental parameters (see Fig.~\ref{Fig1}f-h). The activity cessation time $\tau_\text{off}$ in a range from 10 ms up to 5 minutes has been probed and resulted in no significant effect on the reversal probability upon the re-energizing the system. This observation allows us to exclude the effects caused by remnant hydrodynamic flows or polarization.  When $\tau_\text{off}$ is comparable to $\tau_\text{MW}$ and $\tau_{\nu}$, the vortex does not terminate completely and its subsequent reversal is not triggered. Persistence of the rotation at small  $\tau_\text{off}$ is supported by the incomplete depolarization of the particles and the inertia of a macroscopic hydrodynamic flow.

The robustness of the  chiral state reversal  suggests the presence of a dynamic state memory in the ensemble. The information that is preserved long after the termination of activity, can be stored only in the particle positional arrangements of the ensemble.

The reversal probability depends on the system diameter, $D$, and particle area fraction $\phi$ (see Fig.~\ref{Fig1}g, h). For small diameters the persistence length of roller trajectories is reduced by the high curvature of the walls and the vortex formation probability decreases. The chiral state reversal is robust in a wide range of the particle number densities ($0.04<\phi<0.12$). For dilute area fractions ($\phi<0.03$), the inter-particle interactions are weak and motion of the rollers is uncorrelated. At higher area fractions ($\phi>0.19$), while the vortex formation probability is  high, the reversal probability decreases. The repulsions between the particles at high particle number density reduce the degree of density variations in the system, leading to a decrease in the ``quality'' of the stored information in asymmetric particle arrangements against the background noise.

\begin{figure*}[!t]
\centering
\includegraphics[width=1.0\linewidth]{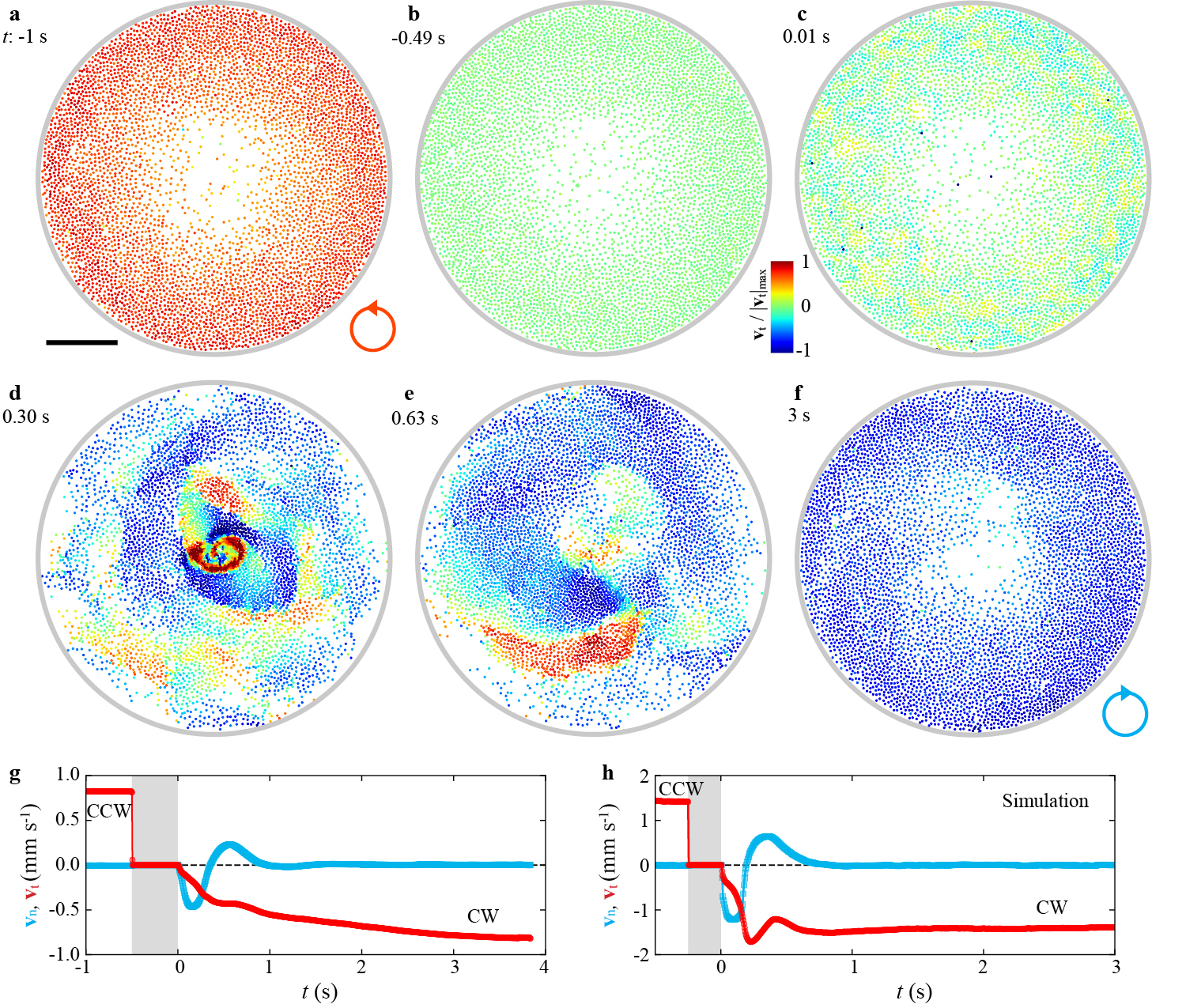}
\caption{
\textbf{Rollers dynamics during chirality reversal of the polar state.}
	\textbf{a}-\textbf{f}, Snapshots of a roller vortex undergoing the chiral state reversal.
	The electric field is reapplied at $t=$ 0 s after a short cessation of activity, $\tau_\text{off}$= 0.5 s. Rollers are shown as circles colored according to the magnitude of the tangential velocity $\bf{v}_\text{t}$.
	\textbf{a}, Particles in a stable CCW vortex.
	\textbf{b}, Particles are at rest when the electric field is off.
	\textbf{c}, Rollers ensemble moments after the system is re-energized.
	\textbf{d}, Most of the particles move towards the center of the well. CW (blue) and CCW (red) roller flocks  are formed.
	\textbf{e}, Particles redistribute over the well. The CW rotating flocks start to dominate over those with CCW rotation.
	\textbf{f}, A stable CW vortex is formed.
	The scale bar is 0.2 mm. See also Supplementary Video 2.
	\textbf{g}, Time evolution of the average normal (blue) and tangential (red) velocities during the chiral state reversal.
	\textbf{h}, Same process as in \textbf{g} is calculated in simulations (see Methods and Supplementary Note 1). Gray areas indicate the state with the electric filed off.
}
\label{Fig2}
\end{figure*}

\begin{figure*}
\centering
\includegraphics[width=1.0\linewidth]{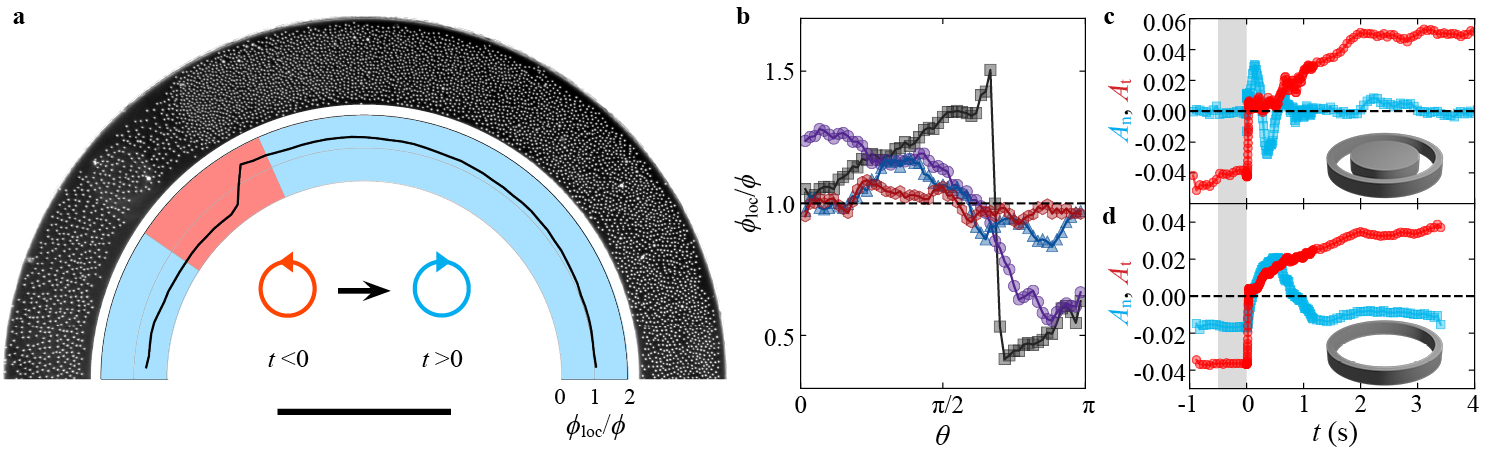}
\caption{
\textbf{Positional order asymmetries in polar active fluids.}
	\textbf{a}, Microscopic image illustrates the spatial distribution of rollers in a circular track (only half of the track is shown) with a width $W=$ 0.25 mm. Particles move CCW and form a stable traveling band along the track. The black curve in the inner ring illustrate the average azimuthal particle density. The background color of the inner ring indicates the direction of initial rollers velocities at the moment when the DC electric field is re-applied. The scale bar is 0.5 mm. See also Supplementary Video 3.
	\textbf{b}, Azimuthal density distribution of rollers in circular tracks with outer diameter $D=2$ mm and the width $W=$ 0.25 mm (black), 0.50 mm (purple), 0.75 mm \SI{} (blue) and in the well (red) as a function of azimuthal angle. Average particles area fraction $\phi = 0.12$. A half of the track (or well) is shown.
	\textbf{c}-\textbf{d}, Evolutions of local order parameters $A_\text{n}$ (blue) and $A_\text{t}$ (red) in a track of the width $W=$ 0.25 mm (\textbf{c}) and a well with $D=$ 2 mm (\textbf{d}) during a vortex reversal. The initial direction of the vortex is CCW. The area fraction $\phi = $  0.12. Oscillations of $A_\text{n}$ at the initial stage of the band formation in a track correspond to the reflections of density bands from the walls. A non-zero value of $A_\text{n}$ reflects the existence of a particle density gradient along the normal direction in the well. Shaded areas indicate field-off time.
}
\label{Fig3}
\end{figure*}

While the observed chiral state reversal is a remarkably robust phenomenon, the formation of a new polar state upon re-activation proceeds through a seemingly chaotic process (see Supplementary Video 2). In a stable vortex illustrated in Fig.~\ref{Fig2}a, all particles initially move CCW with an average tangential velocity of $\textbf{v}_\text{t}$ = 0.82 mm s$^{-1}$ and nearly zero normal velocity.
When the DC electric field is switched off, all particles cease to roll without significant alteration of their relative positions set during the vortex rotation (Fig.~\ref{Fig2}b). After the field is restored, the particles restart motion in seemingly random directions (Fig.~\ref{Fig2}c). Flocks of rollers with spatially correlated velocities form (Fig.~\ref{Fig2}d,e ). The center of the well becomes the densest part of the system in a fraction of a second, and within the next few seconds, particles redistribute into a new stable vortex with a depletion zone in the center (Fig.~\ref{Fig2}f). The process is characterized by the behavior of the average tangential and normal components of the rollers velocities shown in Fig.~\ref{Fig2}g.  At the very first moments upon reactivation, the tangential motion of particles is mostly chaotic, manifested by the emergence of small flocks of rollers moving in both CW and CCW directions. In contrast, the normal component of the average roller velocities sharply increases at re-activation and eventually subside to zero.  While initial fractions of particles moving CCW and CW are approximately the same, the CW fraction steadily grows with time and eventually plateaus with the average tangential velocity reaching $\textbf{v}_\text{t} = -0.82$ mm s$^{-1}$ (equal in magnitude but opposite in the direction to the initial vortex).

\begin{figure*}[!t]
\centering
\includegraphics[width=1.0\linewidth]{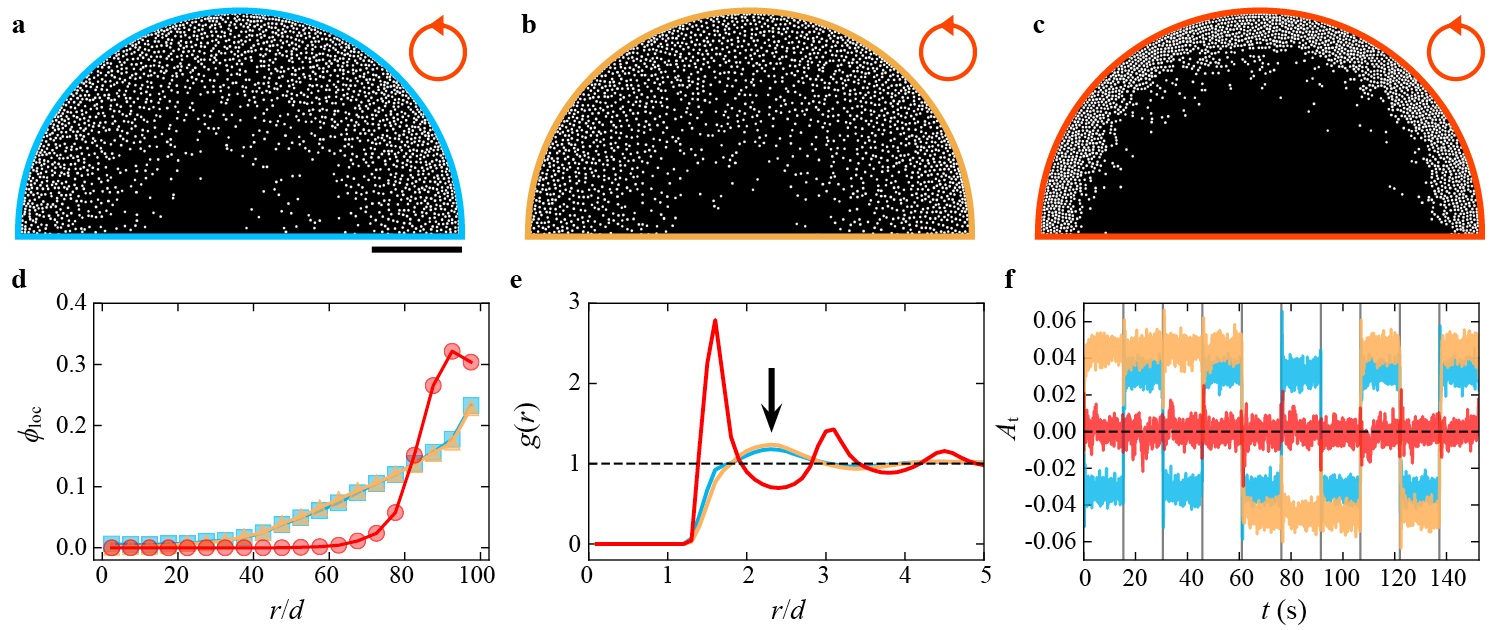}
\caption{
\textbf{Role of microscopic interactions in the chiral state reversal.}
	\textbf{a}-\textbf{c}, Simulation snapshots of stable spontaneous vortices formed under three different scenarios. \textbf{a}, Both electrostatic and hydrodynamic interactions between rollers are included in the model. \textbf{b}, only hydrodynamic interactions drive rollers dynamics. \textbf{c}, only electrostatic interactions contribute to the collective dynamics.  Only half of the wells are shown. The scale bar is 0.2 mm. The particle area fraction $\phi$ = 0.10 and the field strength is $E=$ 2.0 V \SI{}{\micro\meter}$^{-1}$.
	\textbf{d}, Radial variations of local particle density, $\phi_\text{loc}$, in a stable vortex formed with both electrostatic and hydrodynamic interactions (blue), only hydrodynamic (orange), and only electrostatic (red) interactions. $r$ is the distance of particles from the center of the well.
\textbf{e}, Positional pair correlation functions $g(\text{r})$ of particles in stable vortices shown in \textbf{a-c}. The same color coding is used as in \textbf{d}. The black arrow in \textbf{e} points to the identical location of a first peak for scenarios \textbf{a} and \textbf{b}.
\textbf{f}, Local order parameter $A_\text{t}$ calculated for 10 activity cessation cycles. The ensemble does not store any information in the positional order when only electrostatic interactions govern collective dynamics of the rollers (red curve). Local positional order asymmetries encoding the information about the chiral state of the ensemble develop only for scenarios including hydrodynamic interactions (blue, orange).  Nevertheless,  robust reversal is observed only when both hydrodynamic and electrostatic interactions are present(blue).
}
\label{Fig4}
\end{figure*}

The chirality of the collective motion can be similarly reversed in an annulus (see Supplementary Video 3). However, in contrast to the cylindrical confinement, particle traveling bands with a visible local density gradient along the direction of motion are observed in these experiments. The traveling density bands carry the information about the chirality of the polar state. Upon activity termination, it is possible to recover the previous direction of the band motion by a snapshot of the track. In a stable CCW-rotating band  shown in Fig.~\ref{Fig3}a, the local particle density $\phi_\text{loc}$ increases slowly with an azimuth angle toward the front of the band, and then abruptly drops at the frontier. When the system is re-energized, the inter-particle repulsive interactions push rollers away from the high-density regions resulting in a reversal of the band motion.

As the width of the track grows, the tangential density gradient (i.e., the band structure) dissolves and completely vanishes in the case of a well (Fig.~\ref{Fig3}b). Nevertheless, the reversal of the vortex chiral state upon temporal modulation of activity is preserved. It suggests that a structural asymmetry encoded in the local particle positions is now hidden by being redistributed over the whole ensemble. To unveil intrinsic asymmetries in the positional arrangements of rollers in a vortex, we introduce two local order parameters $A_\text{n}$ and $A_\text{t}$.
\begin{equation} \label{A}
A_\text{n,t} =  \langle \mathbf{F}_i  \cdot \mathbf{u}^i_\text{n,t}  \rangle_i.
\end{equation}
Here, $\mathbf{F}_{i}$ is a model repulsive unit force acting on the particle $i$ from the the nearest neighbors (See Supplementary Note 2), and $\mathbf{u}^i_\text{n,t}$ are normal and tangential unit vectors respectively at the position of the particle $i$. $\langle ...\rangle_i$ indicates averaging over all particles. We can simplify the averaging over the neighbors by taking into account only a single nearest particle in Eq. \ref{A}.

Equipped with two order parameters we analyze the reversal process in a circular track and a well. The non-zero value of $A_\text{t}$ in the track during a stable collective motion indicates the density gradient along the tangential direction, see Fig.~\ref{Fig3}c. Simultaneously, $A_\text{n}$ is zero within the noise level during the band motion, indicating the absence of normal particle density gradients in a track.
Both $A_\text{n}$ and $A_\text{t}$ remain nearly constant while the system is in a stable dynamic state. The positive (or negative) value of  $A_\text{t}$ indicates a CW (or CCW) vortex or band.  In a well, the value of $A_\text{t} $ also exhibits stable non-zero values comparable in magnitude to those observed in narrow tracks, even though the density bands are absent (Fig.~\ref{Fig3}d).

To elucidate the physics underlying the chiral state memory formation in the ensemble of active rollers and uncover mechanisms leading to the encoding and subsequent retrieval of the information by the active system we develop a  particle-based model by coupling the Quincke rotation mechanism\cite{quincke1896ueber} with Stokesian Dynamics\cite{Durlofsky1987}  (see Methods and Supplementary Note 1). The simulations correctly capture all phenomenology observed in the roller ensemble upon temporal modulation of the activity (Fig.~\ref{Fig2}h, Supplementary Figure 1 and Supplementary Video 4, 5).

In ensembles of Quincke rollers both electrostatic and hydrodynamic interactions contribute to the velocity alignment processes resulting in a coherent large-scale motion~\cite{bricard2013emergence}. To identify the microscopic mechanisms driving the emergence of a dynamic state memory encoded in the particles positional arrangements, we investigate the behavior of the system when either electrostatic or hydrodynamic interactions are ``turned off'' in the simulations. Without electrostatic interactions, a stable vortex  still forms and has a structure similar to a regular vortex with both types of interactions present (see Fig.~\ref{Fig4}a,b and Supplementary Video 4). The absence of the electrostatic interactions does not affect significantly the overall particle distribution in the vortex  as indicated by a radial particle density distribution (Fig.~\ref{Fig4}d) and pair correlation function, $g(r)$ (Fig.~\ref{Fig4}e). The vortex also preserves the non-zero local order parameter $A_\text{t}$ indicating the presence of the local positional order asymmetries (Fig.~\ref{Fig4}f).

In contrast, the absence of the long-range hydrodynamic interactions results in particle accumulation near the boundary of the well (Fig.~\ref{Fig4}c-d). The vortex forms, however, with most of the particles densely packed near the boundary as indicated by multiple peaks of $g(r)$ (see Fig.~\ref{Fig4}e). Importantly, the order parameter $A_\text{t}$ is at the noise level, see Fig.~\ref{Fig4}f, indicative that the local asymmetries along the tangential direction disappear.
The plots of the order parameter $A_\text{t}$ as the system undergoes activity-inactivity cycles demonstrate that in both cases of truncated interactions no robust vortex reversal is observed, and the probability of the reversal is close to 0.5 suggestive that each time the system randomly selects the chiral state. It implies very specific roles of those two interactions. The hydrodynamic interactions are responsible for the development of asymmetries in the local positional order within the vortex  that effectively encode the information about the chiral state. Electrostatic interactions on the other hand are crucial for de-coding this information in the process of vortex formation. When both ingredients are present the system exhibits remarkable reversibility of its chiral state upon temporal control of the activity by the external field.

To further generalize the results we have constructed a phenomenological model with a bare minimum of ``ingredients'': isotropic short-range repulsions between particles and velocity alignment interactions (see Supplementary Note 3). The model mimics main experimental observations and captures chiral state reversal upon re-energizing of the ensemble with information encoded exclusively in the particle positional order (Supplementary Video 6). It suggests that in general, the presence of short-range isotropic repulsions and velocity alignment interactions is enough to develop the local density asymmetries forming the basis for the ensemble polar state memory.


In this study, we demonstrate that active liquids formed by motile particles are capable of developing memory of their dynamic states and store it in seemingly random positional arrangements of the particles. We show that an ensemble of Quincke rollers stores the information about the globally correlated state via local inter-particle  positional arrangements. The information is preserved long after a complete cessation of activity beyond the Maxwell-Wagner and viscous times. Surprisingly, a relatively weak level of the local arrangement asymmetry is enough to  prescribe the direction of the global vortex motion with high fidelity.

We isolate the role of hydrodynamics as a driving force in the development of the ensemble chiral state memory, and reveal the crucial role of electrostatic repulsive interactions as a main mechanism for the system to access the stored information. The dynamics of the global chiral state reversal involves seemingly chaotic evolution of multiple flocks indicative that the information readout in the system is not instantaneous and relies on multiple inter-particle  interactions. However, the reversal process is robust, and a temporal control of the activity can be exploited to systematically command the subsequent polar states of an active liquid. We envision that with the chirality of the emergent states controlled on-demand, active liquids offer new possibilities for a flow manipulation, transport, and mixing at the microscale.

\bibliographystyle{naturemag}
\bibliography{Refs}

\begin{thebibliography}{10}
\expandafter\ifx\csname url\endcsname\relax
  \def\url#1{\texttt{#1}}\fi
\expandafter\ifx\csname urlprefix\endcsname\relax\def\urlprefix{URL }\fi
\providecommand{\bibinfo}[2]{#2}
\providecommand{\eprint}[2][]{\url{#2}}

\bibitem{vicsek2012collective}
\bibinfo{author}{Vicsek, T.} \& \bibinfo{author}{Zafeiris, A.}
\newblock \bibinfo{title}{Collective motion}.
\newblock \emph{\bibinfo{journal}{Physics Reports}}
  \textbf{\bibinfo{volume}{517}}, \bibinfo{pages}{71--140}
  (\bibinfo{year}{2012}).

\bibitem{marchetti2013hydrodynamics}
\bibinfo{author}{Marchetti, M.~C.} \emph{et~al.}
\newblock \bibinfo{title}{Hydrodynamics of soft active matter}.
\newblock \emph{\bibinfo{journal}{Reviews of Modern Physics}}
  \textbf{\bibinfo{volume}{85}}, \bibinfo{pages}{1143} (\bibinfo{year}{2013}).

\bibitem{aranson2013active}
\bibinfo{author}{Aranson, I.~S.}
\newblock \bibinfo{title}{Active colloids}.
\newblock \emph{\bibinfo{journal}{Physics-Uspekhi}}
  \textbf{\bibinfo{volume}{56}}, \bibinfo{pages}{79} (\bibinfo{year}{2013}).

\bibitem{snezhko2016complex}
\bibinfo{author}{Snezhko, A.}
\newblock \bibinfo{title}{Complex collective dynamics of active torque-driven
  colloids at interfaces}.
\newblock \emph{\bibinfo{journal}{Current Opinion in Colloid \& Interface
  Science}} \textbf{\bibinfo{volume}{21}}, \bibinfo{pages}{65--75}
  (\bibinfo{year}{2016}).

\bibitem{zottl2014hydrodynamics}
\bibinfo{author}{Z{\"o}ttl, A.} \& \bibinfo{author}{Stark, H.}
\newblock \bibinfo{title}{Hydrodynamics determines collective motion and phase
  behavior of active colloids in quasi-two-dimensional confinement}.
\newblock \emph{\bibinfo{journal}{Physical review letters}}
  \textbf{\bibinfo{volume}{112}}, \bibinfo{pages}{118101}
  (\bibinfo{year}{2014}).

\bibitem{wu2017transition}
\bibinfo{author}{Wu, K.-T.} \emph{et~al.}
\newblock \bibinfo{title}{Transition from turbulent to coherent flows in
  confined three-dimensional active fluids}.
\newblock \emph{\bibinfo{journal}{Science}} \textbf{\bibinfo{volume}{355}},
  \bibinfo{pages}{eaal1979} (\bibinfo{year}{2017}).

\bibitem{bricard2015emergent}
\bibinfo{author}{Bricard, A.} \emph{et~al.}
\newblock \bibinfo{title}{Emergent vortices in populations of colloidal
  rollers}.
\newblock \emph{\bibinfo{journal}{Nature communications}}
  \textbf{\bibinfo{volume}{6}}, \bibinfo{pages}{7470} (\bibinfo{year}{2015}).

\bibitem{kaiser2017flocking}
\bibinfo{author}{Kaiser, A.}, \bibinfo{author}{Snezhko, A.} \&
  \bibinfo{author}{Aranson, I.~S.}
\newblock \bibinfo{title}{Flocking ferromagnetic colloids}.
\newblock \emph{\bibinfo{journal}{Science advances}}
  \textbf{\bibinfo{volume}{3}}, \bibinfo{pages}{e1601469}
  (\bibinfo{year}{2017}).

\bibitem{peruani2012collective}
\bibinfo{author}{Peruani, F.} \emph{et~al.}
\newblock \bibinfo{title}{Collective motion and nonequilibrium cluster
  formation in colonies of gliding bacteria}.
\newblock \emph{\bibinfo{journal}{Physical review letters}}
  \textbf{\bibinfo{volume}{108}}, \bibinfo{pages}{098102}
  (\bibinfo{year}{2012}).

\bibitem{schaller2010polar}
\bibinfo{author}{Schaller, V.}, \bibinfo{author}{Weber, C.},
  \bibinfo{author}{Semmrich, C.}, \bibinfo{author}{Frey, E.} \&
  \bibinfo{author}{Bausch, A.~R.}
\newblock \bibinfo{title}{Polar patterns of driven filaments}.
\newblock \emph{\bibinfo{journal}{Nature}} \textbf{\bibinfo{volume}{467}},
  \bibinfo{pages}{73} (\bibinfo{year}{2010}).

\bibitem{sumino2012large}
\bibinfo{author}{Sumino, Y.} \emph{et~al.}
\newblock \bibinfo{title}{Large-scale vortex lattice emerging from collectively
  moving microtubules}.
\newblock \emph{\bibinfo{journal}{Nature}} \textbf{\bibinfo{volume}{483}},
  \bibinfo{pages}{448} (\bibinfo{year}{2012}).

\bibitem{sokolov2012physical}
\bibinfo{author}{Sokolov, A.} \& \bibinfo{author}{Aranson, I.~S.}
\newblock \bibinfo{title}{Physical properties of collective motion in
  suspensions of bacteria}.
\newblock \emph{\bibinfo{journal}{Physical review letters}}
  \textbf{\bibinfo{volume}{109}}, \bibinfo{pages}{248109}
  (\bibinfo{year}{2012}).

\bibitem{bricard2013emergence}
\bibinfo{author}{Bricard, A.}, \bibinfo{author}{Caussin, J.-B.},
  \bibinfo{author}{Desreumaux, N.}, \bibinfo{author}{Dauchot, O.} \&
  \bibinfo{author}{Bartolo, D.}
\newblock \bibinfo{title}{Emergence of macroscopic directed motion in
  populations of motile colloids}.
\newblock \emph{\bibinfo{journal}{Nature}} \textbf{\bibinfo{volume}{503}},
  \bibinfo{pages}{95} (\bibinfo{year}{2013}).

\bibitem{yan2016reconfiguring}
\bibinfo{author}{Yan, J.} \emph{et~al.}
\newblock \bibinfo{title}{Reconfiguring active particles by electrostatic
  imbalance}.
\newblock \emph{\bibinfo{journal}{Nature materials}}
  \textbf{\bibinfo{volume}{15}}, \bibinfo{pages}{1095} (\bibinfo{year}{2016}).

\bibitem{kokot2017active}
\bibinfo{author}{Kokot, G.} \emph{et~al.}
\newblock \bibinfo{title}{Active turbulence in a gas of self-assembled
  spinners}.
\newblock \emph{\bibinfo{journal}{Proceedings of the National Academy of
  Sciences}} \textbf{\bibinfo{volume}{114}}, \bibinfo{pages}{12870--12875}
  (\bibinfo{year}{2017}).

\bibitem{doostmohammadi2018active}
\bibinfo{author}{Doostmohammadi, A.}, \bibinfo{author}{Ign{\'e}s-Mullol, J.},
  \bibinfo{author}{Yeomans, J.~M.} \& \bibinfo{author}{Sagu{\'e}s, F.}
\newblock \bibinfo{title}{Active nematics}.
\newblock \emph{\bibinfo{journal}{Nature communications}}
  \textbf{\bibinfo{volume}{9}}, \bibinfo{pages}{1--13} (\bibinfo{year}{2018}).

\bibitem{zhang2020reconfigurable}
\bibinfo{author}{Zhang, B.}, \bibinfo{author}{Sokolov, A.} \&
  \bibinfo{author}{Snezhko, A.}
\newblock \bibinfo{title}{Reconfigurable emergent patterns in active chiral
  fluids}.
\newblock \emph{\bibinfo{journal}{Nature Communications}}
  \textbf{\bibinfo{volume}{11}}, \bibinfo{pages}{1--9} (\bibinfo{year}{2020}).

\bibitem{martin2013driving}
\bibinfo{author}{Martin, J.~E.} \& \bibinfo{author}{Snezhko, A.}
\newblock \bibinfo{title}{Driving self-assembly and emergent dynamics in
  colloidal suspensions by time-dependent magnetic fields}.
\newblock \emph{\bibinfo{journal}{Reports on Progress in Physics}}
  \textbf{\bibinfo{volume}{76}}, \bibinfo{pages}{126601}
  (\bibinfo{year}{2013}).

\bibitem{driscoll2017unstable}
\bibinfo{author}{Driscoll, M.} \emph{et~al.}
\newblock \bibinfo{title}{Unstable fronts and motile structures formed by
  microrollers}.
\newblock \emph{\bibinfo{journal}{Nature Physics}}
  \textbf{\bibinfo{volume}{13}}, \bibinfo{pages}{375} (\bibinfo{year}{2017}).

\bibitem{weber2013long}
\bibinfo{author}{Weber, C.~A.} \emph{et~al.}
\newblock \bibinfo{title}{Long-range ordering of vibrated polar disks}.
\newblock \emph{\bibinfo{journal}{Physical review letters}}
  \textbf{\bibinfo{volume}{110}}, \bibinfo{pages}{208001}
  (\bibinfo{year}{2013}).

\bibitem{kokot2018manipulation}
\bibinfo{author}{Kokot, G.} \& \bibinfo{author}{Snezhko, A.}
\newblock \bibinfo{title}{Manipulation of emergent vortices in swarms of
  magnetic rollers}.
\newblock \emph{\bibinfo{journal}{Nature communications}}
  \textbf{\bibinfo{volume}{9}}, \bibinfo{pages}{2344} (\bibinfo{year}{2018}).

\bibitem{massana2018active}
\bibinfo{author}{Massana-Cid, H.}, \bibinfo{author}{Codina, J.},
  \bibinfo{author}{Pagonabarraga, I.} \& \bibinfo{author}{Tierno, P.}
\newblock \bibinfo{title}{Active apolar doping determines routes to colloidal
  clusters and gels}.
\newblock \emph{\bibinfo{journal}{Proceedings of the National Academy of
  Sciences}} \textbf{\bibinfo{volume}{115}}, \bibinfo{pages}{10618--10623}
  (\bibinfo{year}{2018}).

\bibitem{han2020reconfigurable}
\bibinfo{author}{Han, K.} \emph{et~al.}
\newblock \bibinfo{title}{Reconfigurable structure and tunable transport in
  synchronized active spinner materials}.
\newblock \emph{\bibinfo{journal}{Science advances}}
  \textbf{\bibinfo{volume}{6}}, \bibinfo{pages}{eaaz8535}
  (\bibinfo{year}{2020}).

\bibitem{palacci2013living}
\bibinfo{author}{Palacci, J.}, \bibinfo{author}{Sacanna, S.},
  \bibinfo{author}{Steinberg, A.~P.}, \bibinfo{author}{Pine, D.~J.} \&
  \bibinfo{author}{Chaikin, P.~M.}
\newblock \bibinfo{title}{Living crystals of light-activated colloidal
  surfers}.
\newblock \emph{\bibinfo{journal}{Science}} \textbf{\bibinfo{volume}{339}},
  \bibinfo{pages}{936--940} (\bibinfo{year}{2013}).

\bibitem{quincke1896ueber}
\bibinfo{author}{Quincke, G.}
\newblock \bibinfo{title}{Ueber rotationen im constanten electrischen felde}.
\newblock \emph{\bibinfo{journal}{Annalen der Physik}}
  \textbf{\bibinfo{volume}{295}}, \bibinfo{pages}{417--486}
  (\bibinfo{year}{1896}).

\bibitem{melcher1969electrohydrodynamics}
\bibinfo{author}{Melcher, J.} \& \bibinfo{author}{Taylor, G.}
\newblock \bibinfo{title}{Electrohydrodynamics: a review of the role of
  interfacial shear stresses}.
\newblock \emph{\bibinfo{journal}{Annual review of fluid mechanics}}
  \textbf{\bibinfo{volume}{1}}, \bibinfo{pages}{111--146}
  (\bibinfo{year}{1969}).

\bibitem{Durlofsky1987}
\bibinfo{author}{Durlofsky, L.}, \bibinfo{author}{Brady, J.~F.} \&
  \bibinfo{author}{Bossis, G.}
\newblock \bibinfo{title}{Dynamic simulation of hydrodynamically interacting
  particles}.
\newblock \emph{\bibinfo{journal}{Journal of Fluid Mechanics}}
  \textbf{\bibinfo{volume}{180}}, \bibinfo{pages}{21--49}
  (\bibinfo{year}{1987}).

\bibitem{PhysRevE.87.043014}
\bibinfo{author}{Das, D.} \& \bibinfo{author}{Saintillan, D.}
\newblock \bibinfo{title}{Electrohydrodynamic interaction of spherical
  particles under quincke rotation}.
\newblock \emph{\bibinfo{journal}{Phys. Rev. E}} \textbf{\bibinfo{volume}{87}},
  \bibinfo{pages}{043014} (\bibinfo{year}{2013}).

\bibitem{PhysRevLett.99.094503}
\bibinfo{author}{Pannacci, N.}, \bibinfo{author}{Lobry, L.} \&
  \bibinfo{author}{Lemaire, E.}
\newblock \bibinfo{title}{How insulating particles increase the conductivity of
  a suspension}.
\newblock \emph{\bibinfo{journal}{Phys. Rev. Lett.}}
  \textbf{\bibinfo{volume}{99}}, \bibinfo{pages}{094503}
  (\bibinfo{year}{2007}).

\bibitem{Jones2005}
\bibinfo{author}{Jones, T.~B.} \& \bibinfo{author}{Jones, T.~B.}
\newblock \emph{\bibinfo{title}{Electromechanics of Particles}}
  (\bibinfo{publisher}{Cambridge University Press}, \bibinfo{year}{2005}).

\bibitem{Sainis2008a}
\bibinfo{author}{Sainis, S.~K.}, \bibinfo{author}{Germain, V.},
  \bibinfo{author}{Mejean, C.~O.} \& \bibinfo{author}{Dufresne, E.~R.}
\newblock \bibinfo{title}{Electrostatic interactions of colloidal particles in
  nonpolar solvents: Role of surface chemistry and charge control agents}.
\newblock \emph{\bibinfo{journal}{Langmuir}} \textbf{\bibinfo{volume}{24}},
  \bibinfo{pages}{1160--1164} (\bibinfo{year}{2008}).

\bibitem{SangtaeKim2005}
\bibinfo{author}{Sangtae~Kim, S. J.~K.}
\newblock \emph{\bibinfo{title}{Microhydrodynamics - Principles and Selected
  Applications}} (\bibinfo{publisher}{Dover Publications, Inc.},
  \bibinfo{year}{2005}).

\bibitem{Brady1988}
\bibinfo{author}{Brady, J.~F.} \& \bibinfo{author}{Bossis, G.}
\newblock \bibinfo{title}{Stokesian dynamics}.
\newblock \emph{\bibinfo{journal}{Annual Review of Fluid Mechanics}}
  \textbf{\bibinfo{volume}{20}}, \bibinfo{pages}{111--157}
  (\bibinfo{year}{1988}).

\bibitem{Rotne1969}
\bibinfo{author}{Rotne, J.} \& \bibinfo{author}{Prager, S.}
\newblock \bibinfo{title}{Variational treatment of hydrodynamic interaction in
  polymers}.
\newblock \emph{\bibinfo{journal}{J. Chem. Phys.}}
  \textbf{\bibinfo{volume}{50}}, \bibinfo{pages}{4831--4837}
  (\bibinfo{year}{1969}).

\bibitem{Wajnryb2013}
\bibinfo{author}{Wajnryb, E.}, \bibinfo{author}{Mizerski, K.~A.},
  \bibinfo{author}{Zuk, P.~J.} \& \bibinfo{author}{Szymczak, P.}
\newblock \bibinfo{title}{Generalization of the rotne-prager-yamakawa mobility
  and shear disturbance tensors}.
\newblock \emph{\bibinfo{journal}{Journal of Fluid Mechanics}}
  \textbf{\bibinfo{volume}{731}}, \bibinfo{pages}{R3} (\bibinfo{year}{2013}).

\bibitem{Blake1971}
\bibinfo{author}{Blake, J.~R.}
\newblock \bibinfo{title}{A note on the image system for a stokeslet in a
  no-slip boundary}.
\newblock \emph{\bibinfo{journal}{Mathematical Proceedings of the Cambridge
  Philosophical Society}} \textbf{\bibinfo{volume}{70}},
  \bibinfo{pages}{303--310} (\bibinfo{year}{1971}).

\bibitem{Swan2007}
\bibinfo{author}{Swan, J.~W.} \& \bibinfo{author}{Brady, J.~F.}
\newblock \bibinfo{title}{Simulation of hydrodynamically interacting particles
  near a no-slip boundary}.
\newblock \emph{\bibinfo{journal}{Physics of Fluids}}
  \textbf{\bibinfo{volume}{19}}, \bibinfo{pages}{113306}
  (\bibinfo{year}{2007}).

\bibitem{BalboaUsabiaga2017}
\bibinfo{author}{Balboa~Usabiaga, F.}, \bibinfo{author}{Delmotte, B.} \&
  \bibinfo{author}{Donev, A.}
\newblock \bibinfo{title}{Brownian dynamics of confined suspensions of active
  microrollers}.
\newblock \emph{\bibinfo{journal}{J. Chem. Phys.}}
  \textbf{\bibinfo{volume}{146}}, \bibinfo{pages}{134104}
  (\bibinfo{year}{2017}).

\bibitem{Ermak1978}
\bibinfo{author}{Ermak, D.~L.} \& \bibinfo{author}{McCammon, J.~A.}
\newblock \bibinfo{title}{Brownian dynamics with hydrodynamic interactions}.
\newblock \emph{\bibinfo{journal}{J. Chem. Phys.}}
  \textbf{\bibinfo{volume}{69}}, \bibinfo{pages}{1352--1360}
  (\bibinfo{year}{1978}).

\bibitem{Kazoe2021}
\bibinfo{author}{Kazoe, Y.} \& \bibinfo{author}{Yoda, M.}
\newblock \bibinfo{title}{Measurements of the near-wall hindered diffusion of
  colloidal particles in the presence of an electric field}.
\newblock \emph{\bibinfo{journal}{Appl. Phys. Lett.}}
  \textbf{\bibinfo{volume}{99}}, \bibinfo{pages}{124104}
  (\bibinfo{year}{2021}).

\end{thebibliography}


\begin{thebibliography}{1}
\expandafter\ifx\csname url\endcsname\relax
  \def\url#1{\texttt{#1}}\fi
\expandafter\ifx\csname urlprefix\endcsname\relax\def\urlprefix{URL }\fi
\providecommand{\bibinfo}[2]{#2}
\providecommand{\eprint}[2][]{\url{#2}}

\bibitem{Swan2007}
\bibinfo{author}{Swan, J.~W.} \& \bibinfo{author}{Brady, J.~F.}
\newblock \bibinfo{title}{Simulation of hydrodynamically interacting particles
  near a no-slip boundary}.
\newblock \emph{\bibinfo{journal}{Physics of Fluids}}
  \textbf{\bibinfo{volume}{19}}, \bibinfo{pages}{113306}
  (\bibinfo{year}{2007}).

\bibitem{SangtaeKim2005}
\bibinfo{author}{Sangtae~Kim, S. J.~K.}
\newblock \emph{\bibinfo{title}{Microhydrodynamics - Principles and Selected
  Applications}} (\bibinfo{publisher}{Dover Publications, Inc.},
  \bibinfo{year}{2005}).

\bibitem{Goldman1967}
\bibinfo{author}{Goldman, A.~J.}, \bibinfo{author}{Cox, R.~G.} \&
  \bibinfo{author}{Brenner, H.}
\newblock \bibinfo{title}{Slow viscous motion of a sphere parallel to a plane
  wall--i motion through a quiescent fluid}.
\newblock \emph{\bibinfo{journal}{Chemical Engineering Science}}
  \textbf{\bibinfo{volume}{22}}, \bibinfo{pages}{637--651}
  (\bibinfo{year}{1967}).

\bibitem{bricard2013emergence}
\bibinfo{author}{Bricard, A.}, \bibinfo{author}{Caussin, J.-B.},
  \bibinfo{author}{Desreumaux, N.}, \bibinfo{author}{Dauchot, O.} \&
  \bibinfo{author}{Bartolo, D.}
\newblock \bibinfo{title}{Emergence of macroscopic directed motion in
  populations of motile colloids}.
\newblock \emph{\bibinfo{journal}{Nature}} \textbf{\bibinfo{volume}{503}},
  \bibinfo{pages}{95} (\bibinfo{year}{2013}).

\bibitem{morin2017distortion}
\bibinfo{author}{Morin, A.}, \bibinfo{author}{Desreumaux, N.},
  \bibinfo{author}{Caussin, J.-B.} \& \bibinfo{author}{Bartolo, D.}
\newblock \bibinfo{title}{Distortion and destruction of colloidal flocks in
  disordered environments}.
\newblock \emph{\bibinfo{journal}{Nature Physics}}
  \textbf{\bibinfo{volume}{13}}, \bibinfo{pages}{63} (\bibinfo{year}{2017}).

\bibitem{grossmann2019particle}
\bibinfo{author}{Gro{\ss}mann, R.}, \bibinfo{author}{Aranson, I.~S.} \&
  \bibinfo{author}{Peruani, F.}
\newblock \bibinfo{title}{A particle-field representation unifies paradigms in
  active matter}.
\newblock \emph{\bibinfo{journal}{arXiv preprint arXiv:1906.00277}}
  (\bibinfo{year}{2019}).

\bibitem{bricard2015emergent}
\bibinfo{author}{Bricard, A.} \emph{et~al.}
\newblock \bibinfo{title}{Emergent vortices in populations of colloidal
  rollers}.
\newblock \emph{\bibinfo{journal}{Nature communications}}
  \textbf{\bibinfo{volume}{6}}, \bibinfo{pages}{7470} (\bibinfo{year}{2015}).

\end{thebibliography}


\subsection*{Acknowledgements}
The research of B. Z., A. Sok., A. S. at Argonne National Laboratory was supported by the U.S. Department of Energy, Office of Science, Basic Energy Sciences, Materials Sciences and Engineering Division. Use of the Center for Nanoscale Materials, an Office of Science user facility, was supported by the U.S. Department of Energy, Office of Science, Office of Basic Energy Sciences, under Contract No. DE-AC02-06CH11357.
H.Y. and M.O.d.l.C. were supported by the Center for Bio-Inspired Energy Science, an Energy Frontier Research Center funded by the US Department of Energy, Office of Science, Basic Energy Sciences under Award DE-SC0000989.


\subsection*{Competing interests}
The authors declare no competing interests.

\clearpage

\subsection*{Methods}
\subsubsection*{Experimental setup}
Polystyrene colloidal particles (G0500, Thermo Scientific) with an average diameter $d$ of 4.8 \SI{}{\micro\meter} are dispersed in a 0.15 mol L$^{-1}$ dioctyl sulfosuccinate sodium (AOT)/hexadecane solution. The colloidal suspension is then injected into a cylindrical chamber made out of a SU-8 cylindrical spacer confined between two parallel ITO-coated glass slides (IT100, Nanocs). A typical height, L, and  inner diameter  of the experimental cell, D, are 45 \SI{}{\micro\meter} and 1 mm, respectively. A uniform DC electric field is introduced by applying a voltage between two ITO-coated glass slides.  The field strength $E$ is kept at 2.7 V \SI{}{\micro\meter}$^{-1}$. The period of the pulse function of the DC field is $T =$ 10 s for $\phi<0.1$ and 20 s for $\phi>0.1$ with the zero-voltage time interval of $0.5$ s and $1$ s respectively.

Viscous time scale $\tau_{\nu} \sim L^{2}/\nu$. Here , $\nu$ is kinematic viscosity of the media. $\tau_{\nu} \simeq$ 1 ms.

Dynamics of the colloidal suspensions is captures by an Olympus IX71 inverted microscope (4$\times$ objective) and a fast-speed camera (IL 5, Fastec Imaging).
To optimize the image processing the frame rates are selected at 120 FPS for particle image velocimetry (PIV) and 850-1000 FPS for particle tracking velocimetry (PTV). PIV, PTV and further data analysis are carried out with a custom-made scripts in Matlab, Python, and Trackpy.


\subsubsection*{Measurements of the local area fraction}
The local area fractions $\phi_\text{loc}$ in Fig.~\ref{Fig3}a-b is measured by dividing the track into sectors with 52.4 \SI{}{\micro\meter} outside arc length. $\phi_\text{loc}$ is the particle area fraction in each sector. In Fig.~\ref{Fig4}d, the well is divided into concentric rings with 25 \SI{}{\micro\meter} thickness, and the area fraction is calculated for each ring.

\subsubsection*{Numerical model}
We developed a particle-based simulation of a population of Quincke rollers by combining the Quincke rotation mechanism\cite{quincke1896ueber}, which enables the active motion of particles, together with Stokesian Dynamics\cite{Durlofsky1987} to account for the many-particles hydrodynamic interactions. In this work, the simulation models $N=4096$ spherical particles of radius $a=2.5$ \SI{}{\micro\meter} with dielectric permittivity $\epsilon_\text{p}=3$ and conductivity $\sigma_\text{p}=0$ immersed in a weakly conducting liquid (AOT/hexadecane) with dielectric permittivity $\epsilon_\text{l}=2$ and conductivity $\sigma_\text{l}=2\times 10^{-8}$ S m$^{-1}$. The viscosity of the fluid is $\eta = 2\ \mathrm{mPa\cdot s}$. A uniform external electric field $\v E_0=E_0\uv z$ ($E_0$ = 2 V  \SI{}{\micro\meter}$^{-1}$) is applied along the z-axis. The particles are confined inside a circular well of diameter $D=1\ \mathrm{mm}$, which corresponds to an area fraction of 10 \%. Because the particles are strongly attracted to the bottom electrode surface, for simplicity, the motion of the particles is constrained at the 2D plane $h$ = 0.1$a$ above the bottom electrode.\\
To leading order of approximation\cite{PhysRevE.87.043014,PhysRevLett.99.094503}, each particle carries an electric dipole moment resulting from the dominant dipolar interfacial charge at each particle-fluid interface and evolves according to the surface charge conservation equation:
\begin{equation}
    \d{\v P^\alpha}{t} = \v \Omega^\alpha \times \left(\v P^\alpha - \chi^\infty \v E^\alpha\right) -\frac{1}{\tau}\left(\v P^\alpha - \chi^0 \v E^\alpha\right)
    \label{eqn-dipole-evolution-multiple}
\end{equation}
where $\v P^\alpha$ and $\v \Omega^\alpha$ are the electric dipole moment and the angular velocity of the $\alpha$-th particle, $\v E^\alpha=\v E_0 + \delta \v E^\alpha$ is the electric field at $\alpha$-th particle's location and $\delta \v E^\alpha$ includes contributions from both the surrounding particles and their image dipoles; $\chi^\infty=4\pi\epsilon_0 a^3 \frac{\epsilon_\text{p}-\epsilon_\text{l}}{\epsilon_\text{p}+2\epsilon_\text{l}}$ and $\chi^0=4\pi\epsilon_0 a^3\frac{\sigma_\text{p}-\sigma_\text{l}}{\sigma_\text{p}+2\sigma_\text{l}}$ are the respective high-frequency and low-frequency polarizability; $\tau=\frac{\epsilon_\text{p}+2\epsilon_\text{l}}{\sigma_\text{p}+2\sigma_\text{l}}$ is the Maxwell-Wagner relaxation time\cite{Jones2005}. Please be noted that eqn. \ref{eqn-dipole-evolution-multiple} does not depend on the fluid flow velocity, which essentially ignores possible electro-osmotic flow effects on the dynamics of surface charge transport.\\
Each Quincke particle interacts with the external electric fields:
\begin{equation}
    \v T^\alpha_{\text{ext}} = \v P^\alpha \times \v E_0
    \label{eqn-interaction-torque}
\end{equation}
and also with its surrounding rollers via the screened dipole-dipole interaction and Coulombic interaction:
\begin{equation}
    U^{\alpha\beta}_{\text{d}}(\v r)=\varepsilon_{\text{d}}\exp{\left(-\kappa_\text{d} r\right)}\left[\frac{1}{r^3}\left(\v P^\alpha \cdot \v P^\beta\right) - \frac{3}{r^5}\left(\v P^\alpha\cdot \v r\right)\left(\v P^\beta \cdot \v r\right)\right]
    \label{eqn-interaction-dipole-dipole}
\end{equation}
\begin{equation}
    U^{\alpha\beta}_{\text{q}}(r)=\varepsilon_\text{q}\frac{\exp{\left(-\kappa_\text{q} r\right)}}{r}
    \label{eqn-interaction-yukawa}
\end{equation}
where $\v r = \v r^\alpha - \v r^\beta$, $r=\abs{\v r}$ and $\v r^\alpha$ is the position of $\alpha$-th particle; $\varepsilon_\text{q}$ and $\kappa_\text{q}^{-1}$ are the interaction strength and the screening length of the Coulombic interaction, which is included to account for steric interactions and possible net charges on each particles. Since the concentration of AOT is 0.15 mol L$^{-1}$, the $\kappa_\text{q}^{-1}$ is estimated\cite{Sainis2008a} as $0.2a$ $(500\  \mathrm{nm})$, which makes the Coulombic interaction highly short-ranged and $\epsilon_\text{q}$ is adjusted to ensure particles do not overlap during simulations. The interaction strength and the screening length of the dipole-dipole interaction are $\varepsilon_\text{d}$ and $\kappa_\text{d}^{-1}$, respectively; $\epsilon_\text{d}=1.0$ and $\kappa_\text{d}^{-1}\approx 5 a \gg \kappa_\text{q}^{-1}$ to ensure that each particle can interact with its neighbors about one particle diameter away.\\
The total force $\v F^\alpha$ and the total torque $\v T^\alpha$ acting on each particles can be obtained by summing over all microscopic interactions. Then, the corresponding motion of each particles are calculated via the mobility matrix formulation based on the configuration of N particles\cite{SangtaeKim2005}:
\begin{equation}
    \begin{pmatrix}
    \v U\\
    \v \Omega
    \end{pmatrix}
    =
    \mathcal{M}
    \begin{pmatrix}
    \v F\\
    \v T
    \end{pmatrix}
    \label{eqn-mobility-matrix}
\end{equation}
where $\v U=\{\v U^1,\cdots,\v U^N\}$ and $\v \Omega=\{\v \Omega^1,\cdots,\v \Omega^N\}$ are the translational and angular velocity vector of N particles, respectively; $\v F=\{\v F^1,\cdots,\v F^N\}$ and $\v T=\{\v T^1,\cdots,\v T^N\}$ are the force and torque vector of N particles, respectively; $\mathcal{M}(\v r^1,\cdots,\v r^N)$ is the grand mobility tensor, which maps the applied forces and torques into the respective translational and angular velocities of each particles.\\
The explicit expressions of the mobility tensor in an unbound medium have been derived via the multipole expansion method for well-separated pair of particles\cite{Brady1988,SangtaeKim2005} or using the Rotne-Prager-Yamakwa (RPY) approximation for overlapping particles\cite{Rotne1969,Wajnryb2013}. Furthermore, additional wall corrections of these mobility tensors have also been derived for the case near an infinite no-slip boundary based on the method of reflection\cite{Blake1971,Swan2007}. In this work, the hydrodynamic interactions between Quincke rollers are modelled with the Rotne-Prager-Blake tensor with wall corrections\cite{driscoll2017unstable,BalboaUsabiaga2017}.\\
Besides the deterministic motions calculated via eqn. \ref{eqn-mobility-matrix}, the stochastic Brownian noise should also be included to account for the thermal fluctuations. In principle, the hydrodynamically-correlated Brownian noise which obeys the fluctuation-dissipation theorem\cite{Ermak1978} should be used, i.e., $\gv \xi = \sqrt{2 k_B T} \mathcal{M}^{1/2} \mathcal{W} + k_B T\div{\mathcal{M}}$, where $k_B$ is the Boltzmann constant, $T$ is the temperature, $\mathcal{W}$ is a vector of independent standard Gaussian noise.
However, for a typical Quincke roller system, the active velocity ($\sim$ 1 mm s$^{-1}$) driven by external electric fields is much large than the diffusive velocity ($\sim$ 0.1 \SI{}{\micro\meter} s$^{-1}$) caused by the thermal energy. The Brownian noise is only important when the roller just starts moving and is negligible once the roller picks up speed. For simplicity, the uncorrelated Brownian noise is used instead:
\begin{equation}
    \gv \xi \approx \sqrt{2 k_B \chi T} \mathcal{M}^{1/2}_{bulk} \mathcal{W}
    \label{eqn-Brownian-noise-uncorrelated}
\end{equation}
where $\mathcal{M}_{bulk}$ is the diagonal grand mobility tensor in an unbound medium and $\chi\approx0.1$ is a factor which accounts for the diffusion hindrance near the boundary\cite{Kazoe2021}.\\
Adding the deterministic hydrodynamic velocities (eqn. \ref{eqn-mobility-matrix}) and the stochastic Brownian velocities (eqn. \ref{eqn-Brownian-noise-uncorrelated}) together, the system evolves according to the following equation of motion:
\begin{equation}
    \d{}{t}
    \begin{pmatrix}
    \v r\\
    \v q
    \end{pmatrix}
    =
    \mathcal{M}
    \begin{pmatrix}
    \v F\\
    \v T
    \end{pmatrix} + \gv \xi
    \label{eqn-eom-multiple}
\end{equation}
where $\v r=\{\v r^1,\cdots,\v r^N\}$ and $\v q=\{\v q^1,\cdots,\v q^N\}$ are the position and orientation vector of N particles, respectively. Solving the above equation of motion (eqn. \ref{eqn-eom-multiple}) simultaneously with the time evolution equation of the electric dipole moment (eqn.\ref{eqn-dipole-evolution-multiple}) enables the dynamic simulations of a population of Quincke rollers.


\end{document}



\maketitle

\clearpage

\section*{Supplementary Note 1: Implementation details of the quantitative particle-based model}
\subsection*{1.1. Time integrators}

In this work, the equation of motion (eqn. 8 in the main text) is solved by a second order Adams-Bashforth method with time step $\Delta t$, i.e.,
\begin{equation}
    \v r^\alpha(t+\Delta t) = \v r^\alpha(t)+\Delta t \left(\frac{3}{2}\v v^\alpha(t)-\frac{1}{2}\v v^\alpha(t-\Delta t)\right)
\end{equation}
where $\v r^\alpha(t)$ and $\v v^\alpha(t)$ are the respective position and velocity vectors of $\alpha$-th particle at time $t$ and $\Delta t$ is generally taken as 0.1 ms - 0.01 ms depending on the system setup to sufficiently resolve the steric interactions and avoid particles overlapping. To improve the stability of the simulations, the maximum displacement in a single time step is limited to $0.1a - 0.025a$ depending on the time step $\Delta t$. The displacement limitation only restricts the large displacements caused by momentary particles overlapping and does not affect the normal dynamics of non-touching particles. The velocity $\v v^\alpha(t)$ includes both the deterministic hydrodynamic velocities and the stochastic Brownian velocities. Since the orientation of particles is irrelevant in this work, the orientations of each particles are not tracked in the simulation.\\
The time evolution equation of electric dipole moment (eqn. 2 in the main text) is an explicit first order ordinary differential equation:
\begin{equation}
    \d{\v P^\alpha}{t}=\v \Omega^\alpha \times \left(\v P^\alpha - \chi^\infty \v E^\alpha\right) -\frac{1}{\tau}\left(\v P^\alpha - \chi^0 \v E^\alpha\right)=f\left[\v P^\alpha(t),\v \Omega^\alpha(t),\v E^\alpha(t)\right]
\end{equation}
and it is solved by a second order Runge-Kutta method with time step $\Delta t^\ast$:
\begin{equation}
\begin{aligned}
    \v P^\alpha (t+\Delta t^\ast) &= \v P^\alpha(t)+\frac{1}{2}\Delta t^\ast\left(k_1+k_2\right)\\
    k_1&=f\left[\v P^\alpha(t),\v \Omega^\alpha(t),\v E^\alpha(t)\right]\\
    k_2&=f\left[\v P^\alpha(t) + \Delta t^\ast k_1,\v \Omega^\alpha(t+\Delta t^\ast),\v E^\alpha(t+\Delta t^\ast)\right]
\end{aligned}
\end{equation}
The time step used for evolving the electric dipole moment is set as $\Delta t^\ast = 0.1 \Delta t$ to better resolve the dynamics of the electric dipole moment. To avoid executing expensive computations multiple times during a time step $\Delta t$, the angular velocities and the electric fields are assumed to be constant during the evolution of dipole moment, i.e. $k_2 \approx f\left[\v P^\alpha(t) + \Delta t^\ast k_1,\v \Omega^\alpha(t),\v E^\alpha(t)\right]$.\\

\subsection*{1.2. Parallel schemes}
The most expensive computation is the N particles hydrodynamic interactions given by eqn. 6 in the main text, which involves the multiplication of a $6N\times6N$ grand mobility tensor $\mathcal{M}$ with a $6N$ forces and torques vector $(\v F;\v T)$. The complexity of a brutal-force calculation scales as $O(N^2)$, which makes it difficult to simulate the length scale and time scale comparable to experimental conditions. However, the linearity of the Stokes equation enables the equivalent block-wise form of eqn. 6 in the main text:
\begin{equation}
    \begin{pmatrix}
    \v U^\alpha\\
    \v \Omega^\alpha
    \end{pmatrix}
    =
    \sum_{\beta}
    \begin{pmatrix}
    \ten{M}_{UF}^{\alpha\beta}&\ten{M}_{UT}^{\alpha\beta}\\
    \ten{M}_{\Omega F}^{\alpha\beta}&\ten{M}_{\Omega T}^{\alpha\beta}
    \end{pmatrix}
    \begin{pmatrix}
    \v F^\beta\\
    \v T^\beta
    \end{pmatrix}
    \label{eqn-mobility-matrix-N}
\end{equation}
where $\v U^\alpha$ and $\v \Omega^\alpha$ are the translational and angular velocity of $\alpha$-th particle, respectively; $\v F^\alpha$ and $\v T^\alpha$ are the force and torque of $\alpha$-th particle, respectively; $\ten{M}_{UF}^{\alpha\beta}$, $\ten{M}_{UT}^{\alpha\beta}$, $\ten{M}_{\Omega F}^{\alpha\beta}$ and $\ten{M}_{\Omega T}^{\alpha\beta}$ are the corresponding self-mobility tensors($\alpha=\beta$) or pair-mobility tensors($\alpha\neq\beta$).\\
The block-wise structure of the matrix-vector product makes it suitable for parallel computation in a matrix-free manner. The whole computation can be paralleled for each pair of particles. For each pair of particles, the computation reduces as a product between a $3\times3$ block mobility tensor $\ten{M}^{\alpha\beta}$ and a $3\times1$ force/torque vector $\v F^\beta$/$\v T^\beta$, which can be computed on-the-fly without explicitly constructing the whole grand mobility tensor $\mathcal{M}$. By utilizing the massive parallel capability of the GPU, the computation efficiency is significantly increased comparing with simulation on the CPU, which makes simulations with experimental conditions practical.\\
Besides the parallel computing of hydrodynamic interactions, the simulation also did the computation in parallel when evolving the electric dipole moment ( eqn. 2 in the main text) and computing the microscopic interactions (eqn. 3-5 in the main text). A typical simulation performance of 4096 particles on a Nvidia Tesla v100 GPU is about 50 time steps per second including the overhead of I/O operations.

\subsection*{1.3. Boundary conditions of bottom walls}
The bottom electrode surface introduces two additional boundary conditions to the system:
\begin{enumerate*}
    \item[1)] the no-slip boundary condition for the hydrodynamics and
    \item[2)] the equi-potential boundary condition for the electrostatics.
\end{enumerate*}
Because of the planar geometry of the bottom electrode, both boundary conditions are included systematically in the simulation.\\
Firstly, the bottom electrode surface acts as a non-slippery boundary which modifies the mobility tensors. The most direct consequence of these wall corrections is a non-zero torque-translation coupling, which enables the self-propulsion of Quincke rollers. Depending on the separation distance between the particle and the bottom electrode, the wall corrections of the mobility tensors can be calculated via the method of reflection\cite{Swan2007} or the lubrication theory\cite{SangtaeKim2005,Goldman1967}. In this work, we follow the former approach to include the wall corrections.\\
Secondly, the bottom electrode surface is a conductor, which can be treated as a an equi-potential surface. The surface charges on the particle-fluid interface induce excess charges on the electrode surface. If we only consider this effect to the dipolar level, these induced charges on the electrode can be effectively replaced by an image dipole $\v P^\ast$. By the image charge method, an electric dipole moment $\v P$ positioned at $(x,y,z)$ induces an image dipole $\v P^\ast$ positioned at $(x,y,-z)$ which has reflected in-plane dipole moments and unchanged out-of-plane dipole moments, i.e., $\v P^\ast=(-P_x,-P_y,P_z)$. Here, we assume that the electrode surface is at $z=0$. To include the effect of image charges, the electric field in eqn. 2 in the main text should include contributions from the image dipoles. Besides, the calculation of electric force and electric torque exerted upon the particle also should include the interactions with image dipoles.\\

\subsection*{1.4. Boundary conditions of side walls}
The no-slip boundary condition of a curved wall (a circular wall in this work) is less straightforward to impose properly. As the side walls do not affect the major results in this work, for simplicity, the boundary condition of side walls is implemented by a simple harmonic wall potential:
\begin{equation}
    U_{\text{wall}}(r)=
    \begin{cases}
    0, r<R-d\\
    \frac{1}{2}\varepsilon_\text{w} \frac{\left(r - R + d\right)^2}{d},R-d \leq r<R\\
    \varepsilon_\text{w} (r-R+\frac{d}{2}),r\geq R
    \end{cases}
\end{equation}
where $r$ is the radial position of particles, $R$ is the radius of the circular well and $d$ is the interaction range of wall potential, which is set as one particle diameter, i.e., $d=2a$. This basically imposes a linearly increasing repulsive force when the particles are within one particle diameter distance from the circular wall and experiences a large constant repulsive force if the particles overlap with the side walls. Here, $\varepsilon_\text{w}$ is the strength of wall potential and its value is chose to be large enough such that no particle-wall overlapping occurs during the simulation.\\
Since the particle-wall collisions exhibit a relatively ballistic behavior according to experimental observations, the normal component of electric dipole moment is also reflected when particles collide with the side walls. This reflective boundary condition ensures a gas-like behavior for dilute systems but does not affect the behaviors of dense systems significantly.

\subsection*{1.5. Simulation Results}
\begin{figure}
\centering
\includegraphics[width=1.0\textwidth]{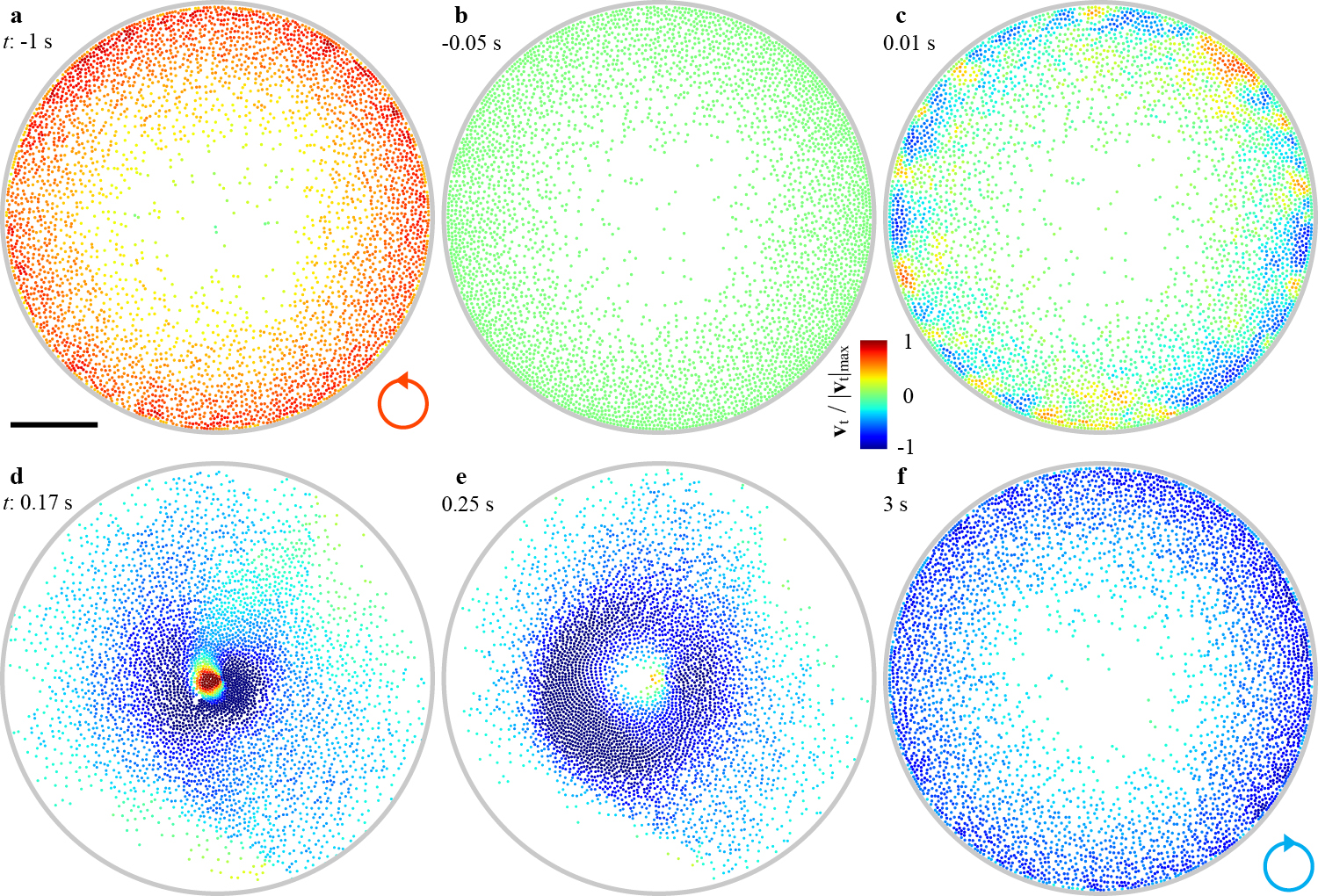}
\caption{
	\textbf{Time evolution of vortex internal structures during a reversal in simulations.}
	\textbf{a}, A stable  CCW roller vortex.
	\textbf{b}, Particles are at rest when the electric field is off.
	\textbf{c}, Motion of the particles immediately after the field is reapplied. Multiple flocks begin to form.
	\textbf{d}, Particles initially move towards the center of the well.
	\textbf{e}, The chirality of the system is now set, rollers redistribute over the well.
	\textbf{f}, A stable roller vortex is assembled. The chirality (CW) is opposite to the initial one.
	The electric field is reapplied at $t=$ 0 s after a short cessation of activity,  ($\tau_\text{off}=$0.25 s). Rollers are shown as circles colored according to the magnitude of the tangential velocity $\bf{v}_\text{t}$.
	The blue and red circles with an arrow indicate the chirality of the vortex. The scale bar is 0.2 mm.
	}
\label{FigS8}
\end{figure}

The quantitative particle-based model successfully capture all essential phenomena observed in the system of Quincke rollers (Supplementary Video 4). The process of the formation of a new polar state via the temporal manipulation of the activity of rollers is very similar to what is observed in experiments, which is confirmed by the time evolution of the particle velocity (Supplementary Figure~\ref{FigS8}, Supplementary Video 5).
To identify the microscopic mechanisms driving the emergence of a dynamic state memory encoded in the particles positional order we investigate the behavior of the system under three different interactions: Scenario \textbf{a}--both electrostatic and hydrodynamic interactions, Scenario \textbf{b}--only hydrodynamic interactions and Scenario \textbf{c}--only electrostatic interactions. The reversal probabilities P are 100 \%, 58 \% and 61 \% calculated from more than 100 cycles for each Scenario \textbf{a}, \textbf{b} and \textbf{c}, respectively.

\clearpage

\section*{Supplementary Note 2: Additional experimental details}

\subsection*{2.1. Vortex reversal in different geometries}
We have carried out additional experiments with rollers in a droplet (no solid perimeter) confined between two ITO-coated glass slides at the same distance of 45 \SI{}{\micro\meter}  (see Supplementary Video 7). While the structure of the 3D flows at the perimeter of the droplets is modified due to a soft fluid/air interface, the process of the vortex self-organization and the probability of the reversal remain unchanged. A series of experiments with two closely located droplets containing Quincke rollers have been conducted. We find that while initial polarities of the two emergent vortices are randomly selected by the system and may be the same or opposite, both vortices consistently reverse their chiral states upon temporal termination of the activity in the system, see Supplementary Video 7.
These experiments exclude the potential influence of the possible asymmetries introduced by external fields. The findings confirm that the polar state reversal  is indeed a result of specific particle configurations developed in a stable vortex that encode the information about the chiral state of the ensemble and preserve it even upon termination of the activity.
The polarity of the collective motion can  also be reversed in an annulus and an elongated track (Supplementary Video 3). However, in contrast to the experiments in the cylindrical confinement, clearly defined particle traveling bands with a well-defined density gradient along the direction of motion are observed, in agreement with previous works \cite{bricard2013emergence, morin2017distortion}.

\subsection*{2.2. Radial velocity profiles and pair correlation functions of rollers in a stable vortex}

\begin{figure}[!b]
\centering
\includegraphics[width=1.0\textwidth]{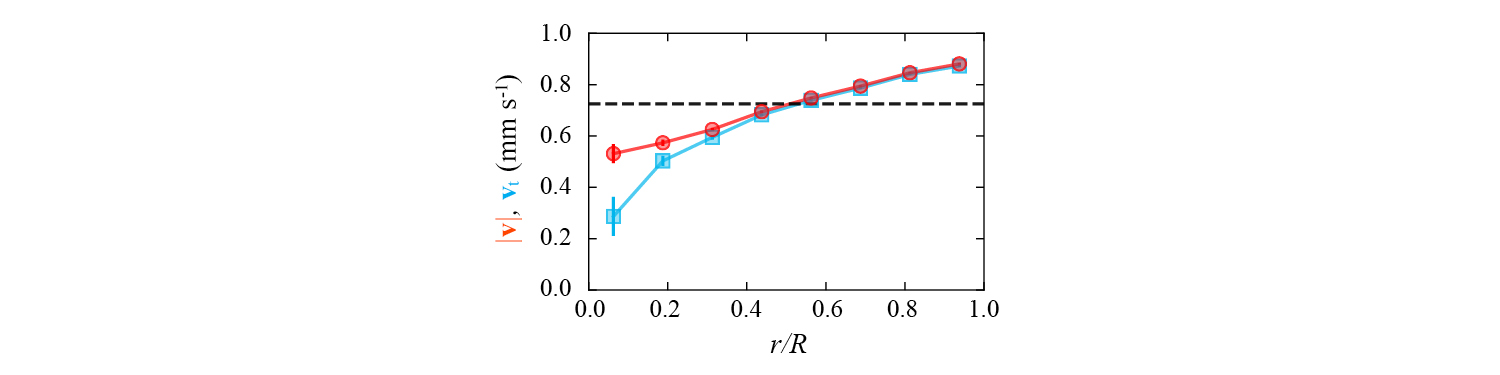}
\caption{
	\textbf{Rollers velocity profile in a stable vortex.}
	Averaged magnitudes of particles instantaneous velocities $|\textbf{v}|$ (red circles) and tangential velocities $\textbf{v}_\text{t}$ (blue squares) are shown as a function of the radial distance $r$ from the center of the vortex.
	The black dash line shows the averaged velocity $v_0$ of an isolated roller measured under the same experimental conditions. The error bars are the standard errors of the corresponding mean. The electric field $E=$ 2.7 V \SI{}{\micro\meter}$^{-1}$ and the particle area fraction $\phi = 0.12$.
}
\label{FigS5}
\end{figure}

The particle tracking velocimetry (PTV) reveals that the tangential velocity $\textbf{v}_\text{t}$ increases with the distance $r$ from the center  (Supplementary Figure~\ref{FigS5}), resembling a solid body rotation, while the normal velocity is close to zero. Such a velocity profile of active liquids allows to minimize the energy dissipation by reducing the shear rate in the system. We also compare velocities of the particles in the stable vortex with the velocity $v_0$ of an individual roller in a very diluted system energized by the same electric field. The rollers near the center of the stable vortex move slower, while the rollers near the perimeter of the well move noticeably faster than $v_0$ (Supplementary Figure~\ref{FigS5}) due to the contribution of macroscopic hydrodynamic flow induced by the vortex.

A typical pair correlation function $g(r)$ of rollers in a stable vortex is shown in Supplementary Figure~\ref{FigS6}. The position of the first peak of $g(r)$--the most probable nearest neighbour distance is $\sim$2.3$d$. The result is very similar to the one from simulations  shown in Fig. 4e in the main text.

\begin{figure}[!t]
\centering
\includegraphics[width=1.0\textwidth]{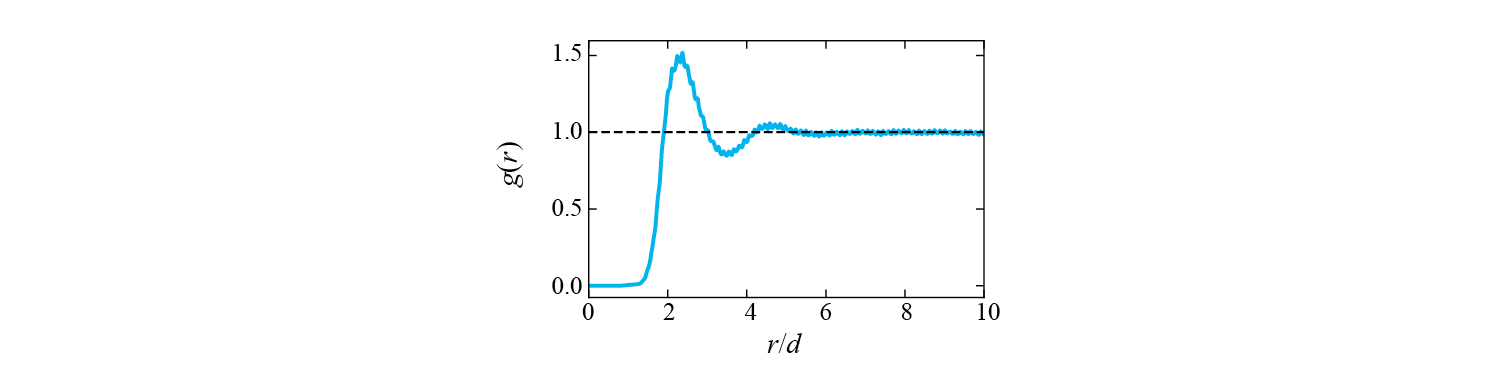}
\caption{
	Pair correlation function $g(r)$ of rollers in a stable vortex.
	The position of the first peak is at 2.3$d$. 	The electric field $E=$ 2.7 V \SI{}{\micro\meter}$^{-1}$, the particle area fraction $\phi = 0.12$.
}
\label{FigS6}
\end{figure}

\begin{figure}[!t]
\centering
\includegraphics[width=1.0\textwidth]{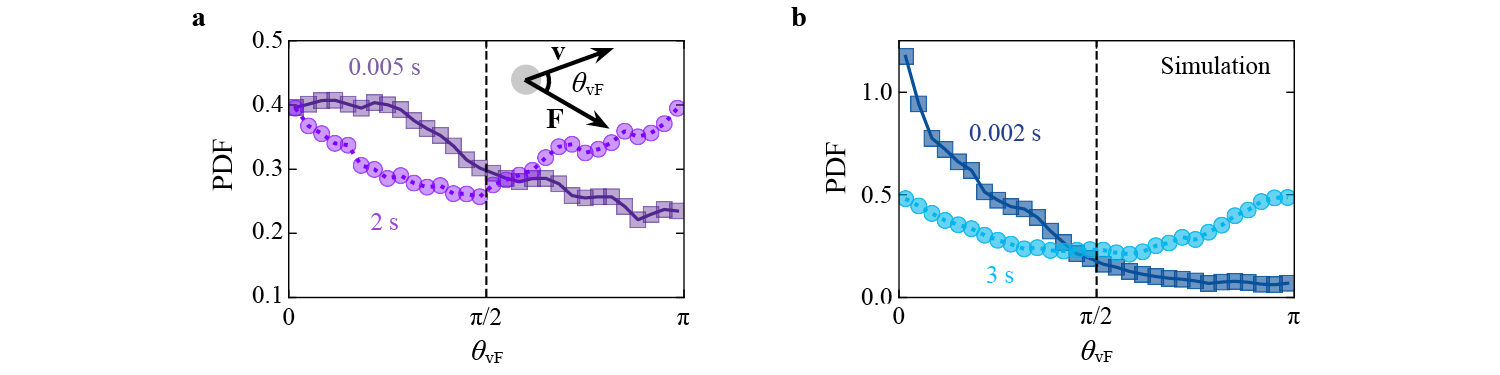}
\caption{
	\textbf{Probability distribution functions of the angle $\theta_\text{vF}$.}
	\textbf{a}, PDFs of $\theta_\text{vF}$ in 0.005 s and 2 s after the system reactivation in experiments.
	\textbf{b}, PDFs of $\theta_\text{vF}$ in 0.002 s and 3 s after the system reactivation in simulations. 	
}
\label{FigS10}
\end{figure}

\subsection*{2.3. Model repulsive forces used to calculate the order parameters.}

In the main text, we define $\mathbf{F}_i$ as a unit repulsive force from the single closest particle in Eq 1. In the general format, $\mathbf{F}_i =  \sum_{j \neq i } \mathbf{f}_{ij} /  | \sum_{j \neq i } \mathbf{f}_{ij} |$.
Here, the unit vector $\mathbf{f}_{ij}$ shows the direction of the force acting on the particle $i$ from the arrangement of neighboring particle $j$ and prescribes the direction of the most probable motion of this particle due to the local environment upon the system reactivation.
We probed several model repulsive potentials to describe inter-particle interactions, including  $\mathbf{f}_{ij}  \sim 1/r_{ij}^n$ and $\mathbf{f}_{ij}  \sim e^{-(r_{ij}/r_0)^n}$, where $r_{ij}=|\mathbf{r}_i-\mathbf{r}_j|$.
The signal-to-noise ratio of the order parameter $A_\text{t}$ increases with $n$.
Generally, the order parameter $A_\text{t}$ accurately indicates the direction of motion when the model potential decays quickly.  For simplicity, we define $\mathbf{F}_i$ as a unit repulsive force from the single closest particle in Eq 1.

\subsection*{2.4. Correlations between initial rollers velocities and electrostatic repulsions}

 Electrostatic repulsions between rollers play an essential role in the initial stage of the vortex formation. The probability distribution functions (PDFs) of $\theta_\text{vF}$--the angle between particle velocity and model unit force calculated from the nearest neighbor particle-- is shown in Supplementary Figure~\ref{FigS10}a. Immediately after the electric field is switched on ($t=$ 0.005 s) corresponding PDF is strongly asymmetric indicative that the majority of the particles initially move close to the expected direction of the repulsive force from a nearest neighbour. The significant coupling between a local density gradient and particles velocities \cite{grossmann2019particle} is only observable at the very early stage of self-organization, while forces created by other active phenomena such as velocity induced alignment, collisions and active noise are not developed. Once the vortex is stable, hydrodynamics takes over and corresponding PDF for $\theta_\text{vF}$ is changed (see PDF curve corresponding to $t=$ 2 s). Similar results  has been observed also in the simulations (see Supplementary Figure~\ref{FigS10}b).

 %


\clearpage

\section*{Supplementary Note 3: Phenomenological model}
\subsection*{3.1. The model}
\begin{figure}
\centering
\includegraphics[width=1.0\linewidth]{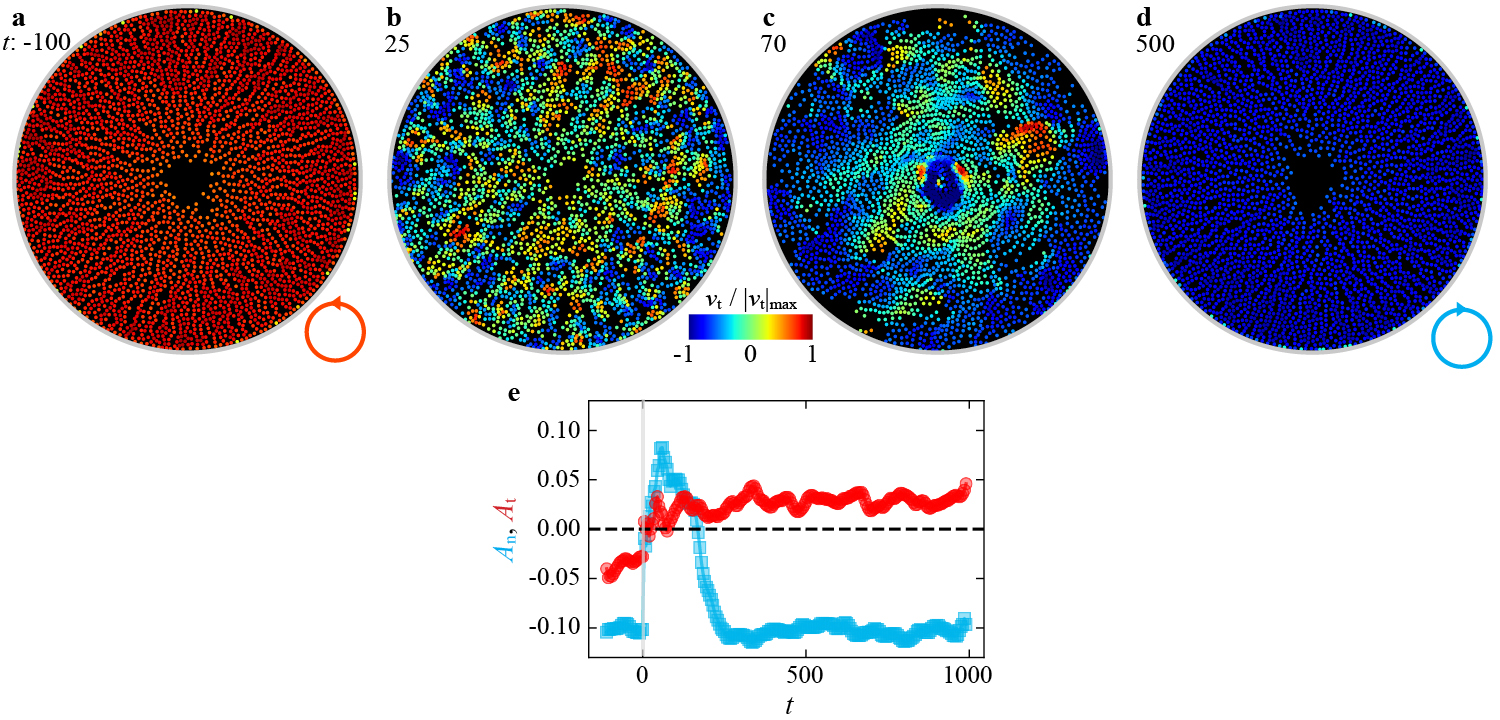}
\caption{
\textbf{Chirality reversal of a vortex under circular confinement as obtained by the phenomenological model.}
	\textbf{a}-\textbf{d}, An example of the vortex reversal from initially CCW state. Colors depict mgnitude of tangential velocities $\textbf{v}_\text{t}$. Blue corresponds to CW, red to CCW.
	\textbf{e}, Evolution of order parameters $A_\text{n}$ (blue squares) and $A_\text{t}$ (red circles) in a well during a vortex reversal. The initial direction of the vortex is CCW.
}
\label{FigS11}
\end{figure}

The reversal of a stable vortex has also been confirmed in the minimalistic phenomenological model. The model is by design instrumental with a bare minimum of ``ingredients" to demonstrate the possibility of a vortex reversal upon re-energizing of the ensemble triggered exclusively by a particle positional order. While the developed phenomenological model does not accurately reflect all physical interactions, the simulations demonstrate a realistic behavior and confirm our main experimental findings. If particles are randomly positioned inside a well or a track, we observe a self-organization of particles into a CW or CCW vortex with equal probability. Upon resetting particle velocities and hydrodynamic flows, which corresponds to a temporal removal of the activity in the system, particles self-organize into a  vortex with reversed polarity (Supplementary Figure~\ref{FigS11}a-d, Supplementary Video 6).

The model is similar to the one previously suggested in Ref.~\cite{bricard2015emergent} with a few simplifications and modifications. The behavior of the system is described in terms of particles coordinates $\mathbf{r}(k)$, velocities $\mathbf{V}(k)$ and hydrodynamic flow $\mathbf{V}_\text{h}(\mathbf{r})$, where $k=1..N$ is a particle index. Here, we need to emphasize that $\mathbf{V}_\text{h}$ is not necessarily a realistic hydrodynamic flow obtained by solving three dimensional Navier-Stokes equation, but a ``tool'' flow introduced to describe far-field particle interactions, mainly velocity alignments. We have tested several types of repulsive potentials, including exponential and polynomial. The qualitative results of the simulation does not change significantly. 
In our model, the dimensionless particle velocity $\mathbf{V}_\text{p}$ is calculated as 
\begin{equation}
\mathbf{V}_\text{p}(k)= \mathbf{V}_0(k) -\nabla \sum_{k' \neq k} U_{k'}(\mathbf{r}(k))-\nabla U_\text{w} (\mathbf{r}(k)) +\alpha_\text{hp} \mathbf{V}_\text{h}(\mathbf{r}(k)) + {\xi},
\label{e1}
\end{equation}
where  $\mathbf{V}_0(k)$ is a unit vector parallel to the orientation of the $k$-th particle, $U_{k'}$ is a repulsive potential created by $k'$-th particle, $U_\text{w} \sim \tanh (\frac{r-R}{d})$ is a repulsive potential of a wall, $\alpha_\text{hp}$ reflects the sensitivity of the particle to the hydrodynamic flow $\mathbf{V}_\text{h}$ created by other particles, and $\xi$ is an uncorrelated white noise.

For better matching of collective dynamics with experimental observations, the hydrodynamic flow $\mathbf{V}_\text{h}$ has a non-negligible (while small) inertia.
\begin{equation}
\dot{\mathbf{V}}_\text{h}(\mathbf{r})=\alpha_1 \sum_k \mathbf{V}_\text{p} e^{-\frac{(\mathbf{r}(k)-\mathbf{r})^2}{\sigma_1^2} } + \alpha_2 \sum \nabla \varphi(k)  - \frac{\mathbf{V}_\text{h}(\mathbf{r})}{\tau_\text{h}} + \mathbf{W}.
\label{e2}
\end{equation}
The first term in Eq \ref{e2} introduces the alignment of particles through hydrodynamic interactions accounted by the third term in Eq \ref{e1}. The meaning of this term can be understood as following. Each particle creates a gaussian-shape flow around it and this ``virtual'' flow aligns other particles in its vicinity. Since each particle creates a complex 3D dimensional flow, we ignored the violation of incomprehensibility condition $\nabla \cdot \mathbf{V}_\text{h} = 0$ in our 2D model introduced by this term. The second term in Eq \ref{e2} describes a potential flow around a moving sphere, where $\varphi(k) \sim 1/r^2 \text{cos}(\theta)$ is velocity potential created by $k$-th particle. According to our simulations this term is not required to visually reproduce the dynamics observed in the experiments and the reversal rate can be 100\% even if $\alpha_2=0$. The third term in Eq \ref{e2} describes dissipation of the flows due to the friction with the walls. The last term $\mathbf{W}$ is introduced to include the influence of a global hydrodynamic vortex created by all particles.
To model this flow we defined $\mathbf{W}$ as
\begin{equation} \label{e3}
\mathbf{W}=\frac{\beta}{N} \sum_k \Big[ \big[\frac{\mathbf{r}(k)}{| \mathbf{r}(k) |} \times \mathbf{V}_\text{p}(k) \big]  \times \frac{\mathbf{r}(k)}{R} \Big].
\end{equation}
The parameter $\beta$ is the strength of particle influence. In our simulations we use a relatively small value of $\beta/ \alpha_1=0.05$. Importantly, the reversal of the vortex and realistic behavior of the system can be observed even if $\beta=0$. The introduction of the small vortical flow $\mathbf{W}$ does not prescribe the direction of the vortex but prevents the formation of dense clusters and stabilizes the global vortex faster.

The evolutions of the order parameters $A_\text{n}$ and $A_\text{t}$ qualitatively match the ones observed in the experiments, see Supplementary Figure~\ref{FigS11}e. A 100$\%{}$ reversal rate has been obtained for a range of simulation parameters, for which particle dynamics in simulations mimics experimental behavior of rollers.

\subsection*{3.2. Parameters of the phenomenological model}

To understand the role of different contributions employed in our model and dependence of the probability of a reversal $P$ we performed a series of simulations covering a wide range of simulation parameters. We analyzed the particle positional order in a stable vortex and probability of a reversal for a range of primary parameters: repulsion radius $r_0$, number of particles $N$ (indicative of a particle number density), and the relaxation time of the hydrodynamic flows $\tau_\text{h}$. The remaining parameters remained fixed: hydrodynamic interaction strength constants $\alpha_1=1$, $\alpha_2=0$, $\alpha_\text{hp}=0.25 $, a hydrodynamic interaction range $\sigma_1=5$, and noise amplitude $\zeta_0=0.02$. The repulsive potential was modeled as $U=e^{-r/r_0}$, where $r$ is the distance from the particle position.

The results of simulations in the well of the diameter $D= 600$ at different values of $r_0$, $N$, and $\tau_\text{h}$ are shown in Supplementary Figure~\ref{FigS12}. The growth of the repulsive radius $r_0$ beyond leads to a drop in the reversal probability, see Supplementary Figure~\ref{FigS12}g. At large $r_0$ particles uniformly fill the well with no depletion zone in the center Supplementary Figure~\ref{FigS12}d. At small $r_0$, the particles are localized along the well's perimeter and form a traveling density band, Supplementary Figure~\ref{FigS12}a. The dependence of the reversal probability on the numbers of particles $N$ qualitatively agrees with the experimental observations, see Supplementary Figure~\ref{FigS12}h. Weakened interaction forces in a very dilute system and a large  noise reduce the reversal probability. In the opposite extreme case, strong repulsive forces and large $N$  reduce density fluctuations. The response of the system to growth of $N$ is qualitatively similar to the increase of the repulsion radius $r_0$.  We observe growth of reversal probability with $\tau_\text{h}$ in our model, see Supplementary Figure~\ref{FigS12}i. It may be contributed to several effects: formation of a larger depletion zone (Supplementary Figure~\ref{FigS12}f), reduction of a noise via effective smoothing of particle motion, and enhanced momentum transfer. 

\begin{figure}
\centering
\includegraphics[width=1.0\textwidth]{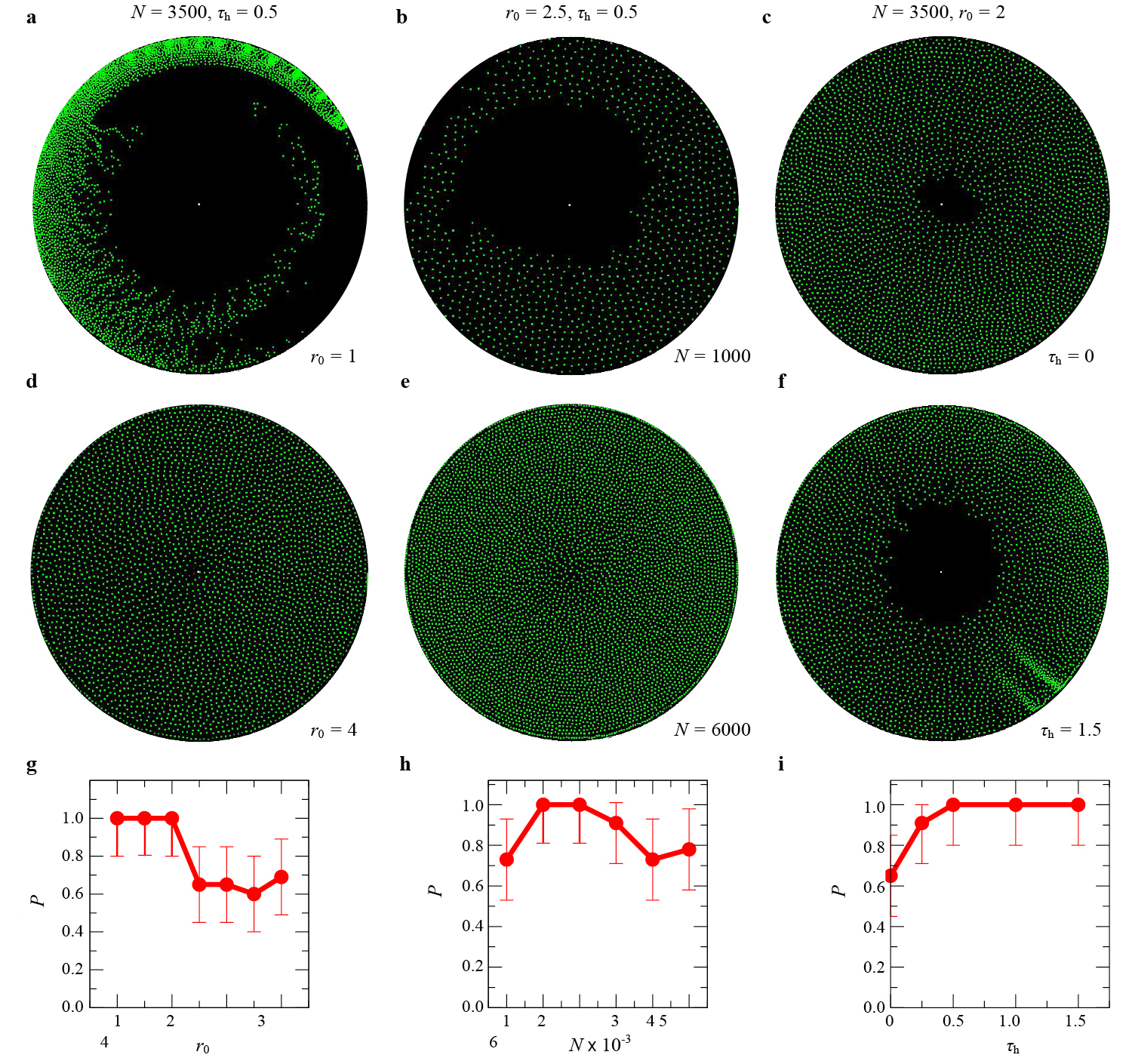}
\caption{
	\textbf{Distribution of particles in a stable vortex and reversal probability for various simulation parameters.}
	\textbf{a}-\textbf{f}, Distribution of particles observed in simulations in a well of diameter $D=$ 600 for different repulsion distance $r_0$ (\textbf{a}, \textbf{d}), number of particles $N$ (\textbf{b}, \textbf{e}), and hydrodynamic relaxation time $\tau_h$ (\textbf{c}, \textbf{f}). The top row (\textbf{a}-\textbf{c}) corresponds to the lowest value of the parameter range presented in (\textbf{g}-\textbf{i}), the bottom row (\textbf{d}-\textbf{f}) to the highest value.
	\textbf{g}-\textbf{i}, The reversal probability $P$ for different value of the repulsion radius $r_0$ (\textbf{g}), the particle quantity $N$ (\textbf{h}), and the relaxation time $\tau_h$ (\textbf{i}). The probability is calculated from 24 simulated cycles of activity.
}
\label{FigS12}
\end{figure}

\clearpage

\bibliographystyle{naturemag}
\bibliography{Refs}